\documentclass[prd,twocolumn,floatfix,amsmath,nofootinbib,amssymb,floatfix]{revtex4}
\usepackage{graphicx,color,dcolumn,booktabs,bm}
\usepackage{longtable,lscape}
\usepackage{pdfpages}
\usepackage{txfonts}
\usepackage{overpic}
\usepackage{amssymb}
\usepackage{makecell}
\usepackage{indentfirst}
\usepackage{feynmf}   
\usepackage{slashed}  
\usepackage{cases}
\usepackage{color}
\usepackage{multirow}
\usepackage{threeparttable}
\usepackage{epstopdf}
\usepackage{enumerate}
\usepackage{subfigure}
\usepackage{diagbox}
\usepackage{graphicx,color,dcolumn,booktabs,bm}
\usepackage[colorlinks,
            citecolor=blue,
            anchorcolor=red,
            menucolor=red,
            linkcolor=red,
            filecolor=red,
            runcolor=red,
            urlcolor=blue,
            frenchlinks=red]{hyperref}

\begin{document}
\title{Systematic analysis of the mass spectra of triply heavy baryons}
\author{Guo-Liang Yu$^{1,2}$}
\email{yuguoliang2011@163.com}
\author{Zhen-Yu Li$^{3}$}
\email{zhenyvli@163.com}
\author{Zhi-Gang Wang$^{1,2}$}
\email{zgwang@aliyun.com}
\author{Ze Zhou$^{1,2}$}

\affiliation{$^1$ Department of Mathematics and Physics, North China
Electric Power University, Baoding 071003, People's Republic of
China\\$^2$ Hebei Key Laboratory of Physics and Energy Technology, North China Electric Power University, Baoding 071003, China\\$^3$ School of Physics and Electronic Science, Guizhou Education University, Guiyang 550018, People's Republic of
China}
\date{\today }

\begin{abstract}
The mass spectra, root mean square (r.m.s.) radii and radial density distributions of $\Omega_{ccb}$ and $\Omega_{bbc}$ baryons are firstly analyzed in the present work. The calculations are carried out in the frame work of relativized quark model, where the baryon is regarded as a real three-quark system. Our results show that the excited energy of charmed-bottom triply baryons are always associated with heavier quark. This means the lowest state of $\Omega_{ccb}$ baryon is dominated by the $\lambda$-mode, however, the dominant orbital excitation for $\Omega_{bbc}$ baryon is $\rho$-mode. In addition, the influence of configuration mixing on mass spectrum, which is induced by different angular momentum assignments, is also analyzed. It shows that energy of the lowest state will be further lowered by this mixing effect. According to this conclusion, we systematically analyze the mass spectra of the ground and excited states($1S\sim4S$, $1P\sim4P$, $1D\sim4D$, $1F\sim4F$ and $1G\sim4G$) of $\Omega_{ccb}$, $\Omega_{bbc}$, $\Omega_{ccc}$ and $\Omega_{bbb}$ baryons. Finally, with the predicated mass spectra, the Regge trajectories of these heavy baryons in the ($J$,$M^{2}$) plane are constructed.
\end{abstract}


\maketitle

\section{Introduction}
In the last two decades, many heavy flavor hadrons such as heavy mesons, single heavy baryons, and hidden-charm tetraquark or pentaquark states were discovered in experiments. Especially, many single heavy baryons have been well confirmed by Belle, BABAR, CLEO and LHCb collaborations \cite{ParticleDataGroup:2024cfk} and the mass spectra of single heavy baryons have become more and more abundance. As for the experimental research about the doubly heavy baryons, experimental physicist also made great breakthrough by the observation of $\Xi_{cc}^{++}$ baryon in 2017 \cite{LHCb:2017iph}. Up to now, only the triply heavy baryons have still not been discovered in the baryon family. Experimentally, higher energy is necessary to produce the triply heavy baryons, and usually, the production rates are not very large \cite{GomshiNobary:2003sf,GomshiNobary:2004mq,GomshiNobary:2005ur,He:2014tga,Zhao:2017gpq}. Especially, it was indicated that the production of triply heavy baryons is extremely difficult in $e^{+}e^{-}$ collision experiments \cite{Baranov:2004er}. The situations are not so pessimistic as predicted by these above literatures. It is optimistic that this ambition may be realized in LHC. In Ref. \cite{Chen:2011mb}, Chen \emph{et} \emph{al}. estimated that $10^{4}-10^{5}$ events of triply heavy baryons with $ccc$ and $ccb$ quark content, could be accumulated for $10$ $fb^{-1}$ integrated luminosity at LHC. In addition, theorists also suggested that people can search for triply heavy baryons in the semi-leptonic and non-leptonic decay processes \cite{Huang:2021jxt,Wang:2022ias,Zhao:2022vfr,Lu:2024bqw}.

Theoretically, investigation of triply heavy baryons is of great interest to physicist, as it provides a good opportunity to understand the strong interactions and basic QCD theory. Up to now, the mass spectra of the triply heavy baryons have been predicted with various methods, such as the bag model \cite{Hasenfratz:1980ka,Bag2}, relativistic or nonrelativistic quark model \cite{Patel:2008mv,Shah:2017jkr,Shah:2018div,Shah:2018bnr,Liu:2019vtx,Yang:2019lsg,Migura:2006ep,Martynenko:2007je,Silves:1996myf,Jia:2006gw,QM0,QM18,Shah:2019jxp,Faustov:2021qqf,Flynn:2011gf,Vijande:2004at}, QCD sum rules \cite{Wang:2011ae,Wang:2019gal,Wang:2020avt,Aliev:2012tt,Azizi:2014jxa,Zhang:2009re,Aliev:2014lxa}, Lattice QCD \cite{Meinel:2010pw,Meinel:2012qz,Padmanath:2013zfa,PACS-CS:2013vie,Vijande:2015faa,Mathur:2018epb,Can:2015exa,Brown:2014ena,Briceno:2012wt}, Regge theory \cite{Wei:2015gsa,Wei:2016jyk,Oudichhya:2023pkg}, potential Non-Relativistic Quantum Chromodynamics(pNRQCD) \cite{Brambilla:2009cd,Llanes-Estrada:2011gwu} and the others \cite{Gutierrez-Guerrero:2019uwa,Brambilla:2005yk,Yin:2019bxe,Qin:2019hgk,Serafin:2018aih}. To our knowledge, most of these studies focused on the mass spectra of ground states, and lower radially or orbitally excited states. The complete mass spectra of triply baryons from ground states to higher radially and orbitally exited states can provide more important information for us to study the properties of these baryons. In addition, the results of different collaborations are not consistent well with each other and need further confirmation by different methods. Thus, it is necessary for us give a systematic analysis of the properties of ground and excited states of triply heavy baryons.

Because triply heavy baryons contain only heavy quarks, they are usually treated as nonrelativistic systems in most literatures. However, investigation of the heavy quark dynamics in heavy quarkonia \cite{Ebert:2002pp,Ebert:2011jc} indicates that the relativistic effects play an important role and should not be neglected in studying the properties of triply heavy baryons. The relativized quark model which was first developed by Godfrey, Capstick and Isgur\cite{GI1,GI2}, is a effective method to achieve this goal. Up to now, it has been widely used to study the properties of the mesons, baryons, and evenly the tetraquark states \cite{LV1,LV2,Wang:2021kfv,Liu:2020lpw,Meng:2023jqk,Yu:2022lak,Yu:2024ljg}. In our previous works \cite{GLY1,Yu:2022lel,ZYL1,Li:2023gbo}, we systematically analyzed the mass spectra of single and doubly heavy baryons with this method. Shortly after the publication of these literatures, several single heavy baryons predicted by us were observed later by LHCb Collaboration. In Ref. \cite{LHCb:2023sxp}, LHCb Collaboration reported two $\Omega_{c}$ resonances with their masses to be $3185.1\pm1.7^{+7.4}_{-0.9}\pm0.2$ and $3327.1\pm1.2^{+0.1}_{-1.3}\pm0.2$ MeV. These values are consistent well with our predicted values for $2S$($\frac{3}{2}^{+}$) and $1D-$wave $\Omega_{c}$ baryons. Besides, another single heavy baryon $\Xi_{b}(6087)$ observed also by LHCb \cite{LHCb:2023zpu} with its mass being $6087\pm0.20\pm0.06\pm0.5$ MeV can be well interpreted as a $1P(\frac{1}{2}^{-})$ or $1P(\frac{3}{2}^{-})$ state by our previous work \cite{ZYL1}.

In the present work, we use the method in Ref. \cite{GLY1} to study the mass spectra and r.m.s. radii of the triply heavy baryons from ground states up to rather high radial and orbital excitations. With the predicted mass spectra, we construct the Regge trajectories in the ($J$,$M^{2}$) plane and determine their Regge slopes and intercepts. The paper is organized as follows. After the introduction, we briefly describe the phenomenological methods adopted in this work in Sec.II. In Sec.III we present our numerical results and discussions about $\Omega_{ccb}$, $\Omega_{bbc}$, $\Omega_{ccc}$ and $\Omega_{bbb}$. In this subsection, the Regge trajectories in the ($J$, $M^{2}$) plane are also constructed. And Sec IV is reserved for our conclusions.

\section{ Phenomenological methods adopted in this work}
\subsection{ Wave function of triply heavy baryon}

\begin{figure}[htbp]
\centering
   \includegraphics[width=10cm]{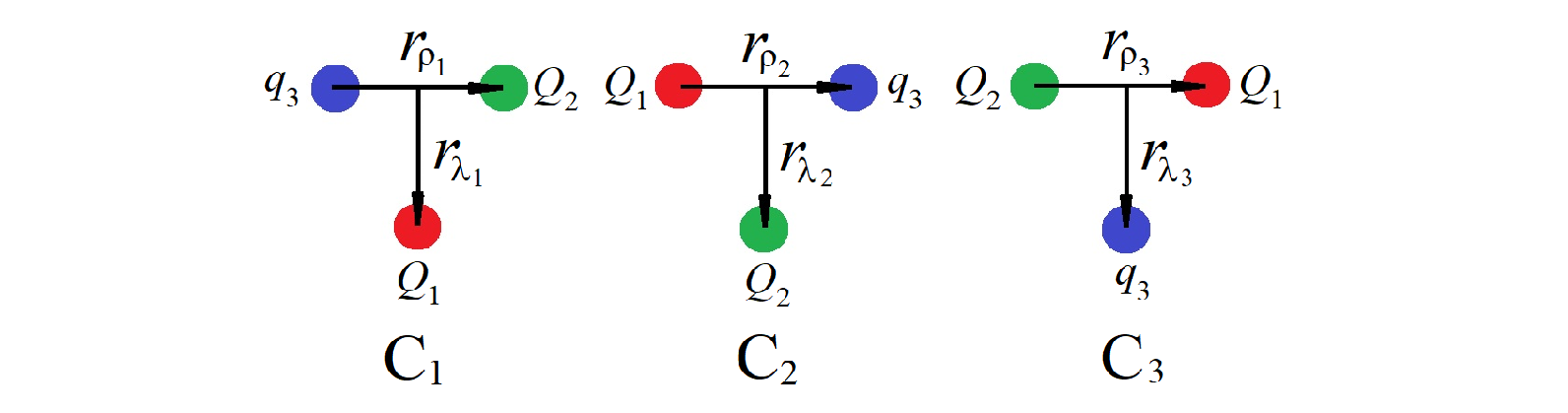}
  \caption{Jacobi coordinates for a three-body system. $Q_{1}$, $Q_{2}$ denote two charmed quarks for $\Omega_{ccb}$, $\Omega_{ccc}$  or two bottom ones for $\Omega_{bbc}$ and $\Omega_{bbb}$. $q_{3}$ represents charmed quark for $\Omega_{ccc}$, $\Omega_{bbc}$ and bottom one for $\Omega_{ccb}$, $\Omega_{bbb}$.}
   \label{figure0}
\end{figure}
The triply heavy baryons are three-body system and their dynamical behavior of inter-quark in this three-body system can be described according to three sets of Jacobi coordinates in Fig. \ref{figure0}. Each set of internal Jacobi coordinate is called a channel ($C$) and is defined as,
\begin{eqnarray}
& \boldsymbol{r}_{\lambda_{i}}=\textbf{r}_{i}-\frac{m_{j}\textbf{r}_{i}+m_{k}\textbf{r}_{k}}{m_{j}+m_{k}} & \\
& \boldsymbol{r}_{\rho_{i}}=\textbf{r}_{j}-\textbf{r}_{k}&
\end{eqnarray}
where $i$, $j$, $k$=1, 2, 3 (or replace their positions in turn). $\mathbf{r}_{i}$ and $m_{i}$ denote the position vector and
the mass of the $i$th quark, respectively.

For $\Omega_{ccb}$ or $\Omega_{bbc}$ baryon, there are two equal quarks in each baryon, thus the mass spectra obtained under $C_{1}$ and $C_{2}$ channels are equivalent with each other. In our previous work, we find a characteristic about the mass spectra of singly and doubly heavy baryons, that their orbital excitations are dominated by heavy quarks. As for charmed-bottom baryons, the bottom quark is much heavier than charmed quark. This implies that these triply heavy baryons may have similar feature to the singly and doubly heavy baryons. It can be seen from $C_{3}$ channel in Fig. \ref{figure0} that the heavy quark degrees of freedom is decoupled from light ones. This channel can properly reflect the characteristic of heavy quark dominance.
Thus, the calculations in this work are performed based on $C_{3}$ channel. Using the transformation
of Jacobi coordinates, we can calculate all the matrix elements in $C_{3}$ channel.
Under this picture, the degree of freedom between two identical quarks is called the $\rho$-mode, while the degree between the center of mass of these two quarks and the other one is called the $\lambda$-mode.

The spatial wave function of a three-body system includes the spin wave function and orbital part, which can be written as,
\begin{align}
\psi_{JM}=\big[\big[[\chi_{1/2}(Q_{1})\chi_{1/2}(Q_{2})]_{s}\Phi_{l_{\rho},l_{\lambda},L}\big]_{j}\chi_{1/2}(q_{3})\big]_{JM}
\end{align}
$\chi_{1/2}$ is the spin wave function of quark and $\textbf{\emph{s}}$ is the total spin of $Q_{1}$ and $Q_{2}$. The orbital wave function is constructed from the wave functions of the two Jacobi coordinates $\rho$ and $\lambda$, and takes the form,
\begin{eqnarray}
\Phi_{l_{\rho},l_{\lambda},L}=\big[\phi_{n_{\rho}l_{\rho}m_{l_{\rho}}}(\boldsymbol{r}_{\rho})\phi_{n_{\lambda}l_{\lambda}m_{l_{\lambda}}}(\boldsymbol{r}_{\lambda})\big]_{L}
\label{eq4}
\end{eqnarray}
The coupling scheme of the spin
and angular momenta is $\textbf{\emph{L}}=\textbf{\emph{l}}_{\rho}+\textbf{\emph{l}}_{\lambda}$, $\textbf{\emph{j}}=\textbf{\emph{s}}+\textbf{\emph{L}}$, $\textbf{\emph{J}}=\textbf{\emph{j}}+\frac{1}{2}$.
In Eq. (\ref{eq4}), $\phi_{nlm_{l}}$ is the Gaussian basis functions \cite{Gaussian1} which can be written as,
\begin{eqnarray}
\phi_{nlm_{l}}(\boldsymbol{r})=N_{nl}r^{l}e^{-\nu_{n}r^{2}}Y_{lm_{l}}(\hat{\boldsymbol{r}}), \quad n=1\sim n_{max}
\label{gauss}
\end{eqnarray}
with
\begin{eqnarray}
N_{nl}=\sqrt{\frac{2^{l+2}(2\nu_{n})^{l+3/2}}{\sqrt{\pi}(2l+1)!!}}
\end{eqnarray}
\begin{eqnarray}
\nu_{n}=\frac{1}{r_{n}^{2}}, \quad r_{n}=r_{a}\Big[\frac{r_{amax}}{r_{a}}\Big]^{\frac{n-1}{n_{max}-1}}
\end{eqnarray}
$n_{max}$ is the maximum number of the Gaussian basis functions, $r_{a}$ and $r_{amax}$ are the Gaussian range parameters. In different studies, people employed different values for these parameters \cite{Gaussian2,Gaussian3}. It is indicated by our previous studies \cite{GLY1,Yu:2022lel} that the results show well stability and convergence with the parameters being taken as $r_{a}$=0.18 fm, $r_{amax}$=15 fm and $n_{max}=10$.

For a three-body system, the calculations of the Hamiltonian matrix elements is very laborious with Gaussian basis functions.
Thus, the Gaussian basis function of Eq. (\ref{gauss}) is substituted by the following infinitesimally-shifted Gaussian (ISG) basis functions \cite{Gaussian2,Gaussian3},
\begin{align}
&\phi_{nlm_{l}}(\boldsymbol{r})=N_{nl}\lim_{\varepsilon\rightarrow 0}\frac{1}{(\nu_{n}\varepsilon)^{l}}\sum_{k=1}^{k_{max}}C_{lm_{l},k}e^{-\nu_{n}(\textbf{r}-\varepsilon \textbf{D}_{lm_{l},k})^{2}}
\end{align}
where $\varepsilon$ is the shifted distance of the Gaussian basis. Taking the limit $\varepsilon\rightarrow 0$ is to be carried out after the Hamiltonian matrix elements have been calculated analytically.
For more details about the ISG basis functions, one can consults our previous work \cite{GLY1}.

For a definite state of a baryon, its full wave function can be expressed as the direct product of color wave function, flavor wave function and the spatial wave function,
\begin{eqnarray}
\Psi_{full}^{JM}=\phi_{\mathrm{color}}\otimes \phi_{\mathrm{flavor}}\otimes\Psi_{JM}(\boldsymbol{r}_{\rho},\boldsymbol{r}_{\lambda})
\end{eqnarray}
with
\begin{align}
\Psi_{JM}(\boldsymbol{r}_{\rho},\boldsymbol{r}_{\lambda})=\sum_{\kappa}C_{\kappa}\psi_{JM}
\end{align}
where $C_{\kappa}$ is expansion coefficients, and $\kappa$ denotes the quantum numbers $\{$$n_{\rho}$, $n_{\lambda}$, $l_{\rho}$, $l_{\lambda}$, $\cdots$ $j$ $\}$.

For these two equal quarks ($Q_{1}Q_{2}$) in $\Omega_{ccb}$ or $\Omega_{bbc}$, their flavor wave function and color function are symmetric and antisymmetric, respectively. The total wave function must be antisymmetric, thus the spatial part should always be symmetric. For this double quark system ($Q_{1}Q_{2}$) in the triply baryon, its spin wave function is either antisymmetric singlet($s=0$) or symmetric triplet($s=1$). To satisfy the symmetry requirements of spatial part, the orbital part must also be antisymmetric for $s=0$ or symmetric for $s=1$. Thus, the total spin $s$ and orbital quantum number $l_{\rho}$ of double quark system should satisfy the condition $(-1)^{s+l_{\rho}} = -1$. As for the $\Omega_{ccc}$ and $\Omega_{bbb}$ baryons, in order to fulfill the Pauli principle, there is no S-wave bound state with the total spin and parity $J^{P}=\frac{1}{2}^{+}$.

\subsection{ The relativized quark model}

In this subsection, we will discuss the Hamiltonian of relativized quark model. Under this theoretical framework, the Hamiltonian for a triply heavy baryon can be written as \cite{GI1,GI2},
\begin{align}\label{Hami}
H=\sum_{i=1}^{3}(p_{i}^{2}+m_{i}^{2})^{1/2}+\sum_{i<j}H_{ij}^{\mathrm{conf}}+\sum_{i<j}H_{ij}^{\mathrm{hyp}}+\sum_{i<j}H_{ij}^{\mathrm{so}}
\end{align}
The first term is called relativistic kinetic energy term, and $H^{\mathrm{conf}}_{ij}$ is the spin-independent potential which is composed by a linear confining potential $S(r_{ij})$ and a one-gluon
exchange potential $\widetilde{G^{\prime}}(r_{ij})$,
\begin{eqnarray}
H^{\mathrm{conf}}_{ij}=S(r_{ij})+\widetilde{G^{\prime}}(r_{ij})
\end{eqnarray}
They can be expressed as,
\begin{eqnarray}\label{Sprime}
& S(r_{ij})=-\frac{3}{4}\textbf{\emph{F}}_{i}\cdot\textbf{\emph{F}}_{j}\Bigg\{b r_{ij}\bigg[\frac{e^{-\sigma_{ij}^{2}r_{ij}^{2}}}{\sqrt{\pi}\sigma_{ij} r_{ij}}
+\bigg(1+\frac{1}{2\sigma_{ij}^{2}r_{ij}^{2}}\bigg)\notag \\ & \times\frac{2}{\sqrt{\pi}} \int^{\sigma_{ij} r_{ij}}_{0}e^{-x^{2}}dx\bigg]+c\Bigg\}
\end{eqnarray}
and
\begin{eqnarray}\label{Gprime}
\widetilde{G^{\prime}}(r_{ij})=\Big(1+\frac{p^{2}_{ij}}{E_{i}E_{j}}\Big)^{\frac{1}{2}}G(r_{ij})\Big(1+\frac{p^{2}_{ij}}{E_{i}E_{j}}\Big)^{\frac{1}{2}}
\end{eqnarray}
with \begin{align}
&\sigma_{ij}=\sqrt{s^{2}\Big[\frac{2m_{i}m_{j}}{m_{i}+m_{j}}\Big]^{2}+\sigma_{0}^{2}\Big[\frac{1}{2}\big(\frac{4m_{i}m_{j}}{(m_{i}+m_{j})^{2}}\big)^{4}+\frac{1}{2}\Big]} &&
\end{align}
In Eq. (\ref{Gprime}), $G(r_{ij})$ is the one-gluon-exchange propagator and it can be expressed as,
\begin{eqnarray}\label{Gww}
G(r_{ij})=\textbf{\emph{F}}_{i}\cdot\textbf{\emph{F}}_{j}\mathop{\sum}\limits_{k=1}^{3}\frac{2\alpha_{k}}{3\sqrt{\pi}r_{ij}}\int^{\tau_{k}r_{ij}}_{0}e^{-x^{2}}dx
\end{eqnarray}
with $\tau_{k}=\frac{1}{\sqrt{\frac{1}{\sigma_{ij}^{2}}+\frac{1}{\gamma_{k}^{2}}}}$.

In Eqs. (\ref{Sprime}) and (\ref{Gww}), $\textbf{\emph{F}}_{i}\cdot\textbf{\emph{F}}_{j}$ stands for the color matrix and $F_{n}$ reads,
\begin{equation}
F_{n}=\left\{
      \begin{array}{l}
       \frac{\lambda_{n}}{2} \quad \mathrm{for} \, \mathrm{quarks}, \\
        -\frac{\lambda_{n}^{*}}{2} \quad    \mathrm{for} \, \mathrm{antiquarks} \\
      \end{array}
      \right.
\end{equation}
with $n=1,2\cdots8$.

In Eq. (\ref{Hami}), $H^{\mathrm{hyp}}$ is the color-hyperfine interaction and it is composed by a tensor term $H^{\mathrm{tensor}}$ and a contact interaction $H^{\mathrm{c}}$, where
\begin{eqnarray}\label{Ht}
H^{\mathrm{tensor}}_{ij}&&=-\Big(\frac{\textbf{S}_{i}\cdot \textbf{r}_{ij}\textbf{S}_{j}\cdot \textbf{r}_{ij}/r_{ij}^{2}-\frac{1}{3}\textbf{S}_{i}\cdot\textbf{S}_{j}}{m_{i}m_{j}}\Big) \notag \\
&& \times\Big(\frac{\partial^{2}}{\partial r_{ij}^{2}}-\frac{1}{r_{ij}}\frac{\partial}{\partial r_{ij}}\Big)\widetilde{G}_{ij}^{\mathrm{t}}
\end{eqnarray}
and
\begin{eqnarray}\label{Hc}
H^{\mathrm{c}}_{ij}=\frac{2\textbf{S}_{i}\cdot\textbf{S}_{j}}{3m_{i}m_{j}}\bigtriangledown^{2}\widetilde{G}_{ij}^{\mathrm{c}}
\end{eqnarray}
The last term in Hamiltonian is the spin-orbit interaction which can also be divided into two parts $H^{\mathrm{so(v)}}$ and $H^{\mathrm{so(s)}}$. These two interactions can be written as,
\begin{eqnarray}\label{Hsov}
 H^{\mathrm{so(v)}}_{ij}&&=\frac{\textbf{S}_{i}\cdot \textbf{L}_{ij}}{2m_{i}^{2}r_{ij}}\frac{\partial \widetilde{G}^{\mathrm{so(v)}}_{ii}}{\partial r_{ij}}+\frac{\textbf{S}_{j}\cdot \textbf{L}_{ij}}{2m_{j}^{2}r_{ij}}\frac{\partial \widetilde{G}^{\mathrm{so(v)}}_{jj}}{\partial r_{ij}}\notag \\
&& +\frac{(\textbf{S}_{i}+\textbf{S}_{j})\cdot \textbf{L}_{ij}}{m_{i}m_{j}r_{ij}}\frac{1}{r_{ij}}\frac{\partial \widetilde{G}^{\mathrm{so(v)}}_{ij}}{\partial r_{ij}}
\end{eqnarray}
and
\begin{eqnarray}\label{Hsos}
H^{\mathrm{so(s)}}_{ij}=-\frac{\textbf{S}_{i}\cdot \textbf{L}_{ij}}{2m_{i}^{2}r_{ij}}\frac{\partial \widetilde{S}^{\mathrm{so(s)}}_{ii}}{\partial r_{ij}}-\frac{\textbf{S}_{j}\cdot \textbf{L}_{ij}}{2m_{j}^{2}r_{ij}}\frac{\partial \widetilde{S}^{\mathrm{so(s)}}_{jj}}{\partial r_{ij}}
\end{eqnarray}
In Eqs. (\ref{Ht})-(\ref{Hsos}), $\widetilde{G}^{\mathrm{t}}_{ij}$, $\widetilde{G}^{\mathrm{c}}_{ij}$, $\widetilde{G}^{\mathrm{so(v)}}_{ij}$ and $\widetilde{S}^{\mathrm{so(s)}}_{ii}$ are achieved from $G(r_{ij})$ and $S(r_{ij})$ by introducing momentum-dependent factors,
\begin{align}
\widetilde{G}^{\mathrm{t}}_{ij}=\Big(\frac{m_{i}m_{j}}{E_{i}E_{j}}\Big)^{\frac{1}{2}+\epsilon_{\mathrm{t}}}G(r_{ij})\Big(\frac{m_{i}m_{j}}{E_{i}E_{j}}\Big)^{\frac{1}{2}+\epsilon_{\mathrm{t}}}
\end{align}
\begin{align}
\widetilde{G}^{\mathrm{c}}_{ij}=\Big(\frac{m_{i}m_{j}}{E_{i}E_{j}}\Big)^{\frac{1}{2}+\epsilon_{\mathrm{c}}}G(r_{ij})\Big(\frac{m_{i}m_{j}}{E_{i}E_{j}}\Big)^{\frac{1}{2}+\epsilon_{\mathrm{c}}}
\end{align}
\begin{align}
\widetilde{G}^{\mathrm{so(v)}}_{ij}=\Big(\frac{m_{i}m_{j}}{E_{i}E_{j}}\Big)^{\frac{1}{2}+\epsilon_{\mathrm{so(v)}}}G(r_{ij})\Big(\frac{m_{i}m_{j}}{E_{i}E_{j}}\Big)^{\frac{1}{2}+\epsilon_{\mathrm{so(v)}}}
\end{align}
\begin{align}
\widetilde{S}^{\mathrm{so(s)}}_{ii}=\Big(\frac{m_{i}^{2}}{E_{i}^{2}}\Big)^{\frac{1}{2}+\epsilon_{\mathrm{so(s)}}}S(r_{ij})\Big(\frac{m_{i}^{2}}{E_{i}^{2}}\Big)^{\frac{1}{2}+\epsilon_{\mathrm{so(s)}}}
\end{align}
with $E_{i}=\sqrt{m_{i}^{2}+p_{ij}^{2}}$, and $\epsilon_{\mathrm{t}}$, $\epsilon_{\mathrm{c}}$, $\epsilon_{\mathrm{so(v)}}$ and $\epsilon_{\mathrm{so(s)}}$ are free parameters which take the same values with those in Ref. \cite{GLY1}. The $p_{ij}$ is the magnitude of the momentum of either of the quarks in the $ij$ center-of-mass frame.

With the Hamiltonian of Eq. (\ref{Hami}), all of the matrix elements can be evaluated, and the mass spectra can be obtained by solving the generalized eigenvalue problem,
\begin{align}
\sum_{j=1}^{n_{max}^{2}}\Big(H_{ij}-EN_{ij}\Big)C_{j}=0, \quad (i=1-n_{max}^{2})
\end{align}
$C_{j}$ is the coefficient of eigenvector, and $N_{ij}$ is the overlap matrix elements of the Gaussian functions, which can be expressed as,
\begin{align}
\notag
&N_{ij}\equiv \langle\phi_{n_{\rho_{a}}l_{\rho_{a}}m_{l_{\rho_{a}}}}|
\phi_{n_{\rho_{b}}l_{\rho_{b}}m_{l_{\rho_{b}}}}\rangle \notag \times\langle\phi_{n_{\lambda_{a}}l_{\lambda_{a}}m_{l_{\lambda_{a}}}}
|\phi_{n_{\lambda_{b}}l_{\lambda_{b}}m_{l_{\lambda_{b}}}}\rangle \notag \\
&=\Big(\frac{2\sqrt{\nu_{n_{\rho_{a}}}\nu_{n_{\rho_{b}}}}}{\nu_{n_{\rho_{a}}}+\nu_{n_{\rho_{b}}}}\Big)^{l_{\rho_{a}}+3/2}\times
\Big(\frac{2\sqrt{\nu_{n_{\lambda_{a}}}\nu_{n_{\lambda_{b}}}}}{\nu_{n_{\lambda_{a}}}+\nu_{n_{\lambda_{b}}}}\Big)^{l_{\lambda_{a}}+3/2} &
\end{align}

\section{Numerical results and discussions}

\subsection{The orbital excitations of $\Omega_{ccb}$ and $\Omega_{bbc}$ baryons}\label{IIIA}

\begin{figure}[htbp]
  \centering
   \subfigure[]{
   \begin{minipage}{6.5cm}
   \centering
   \includegraphics[width=6.5cm]{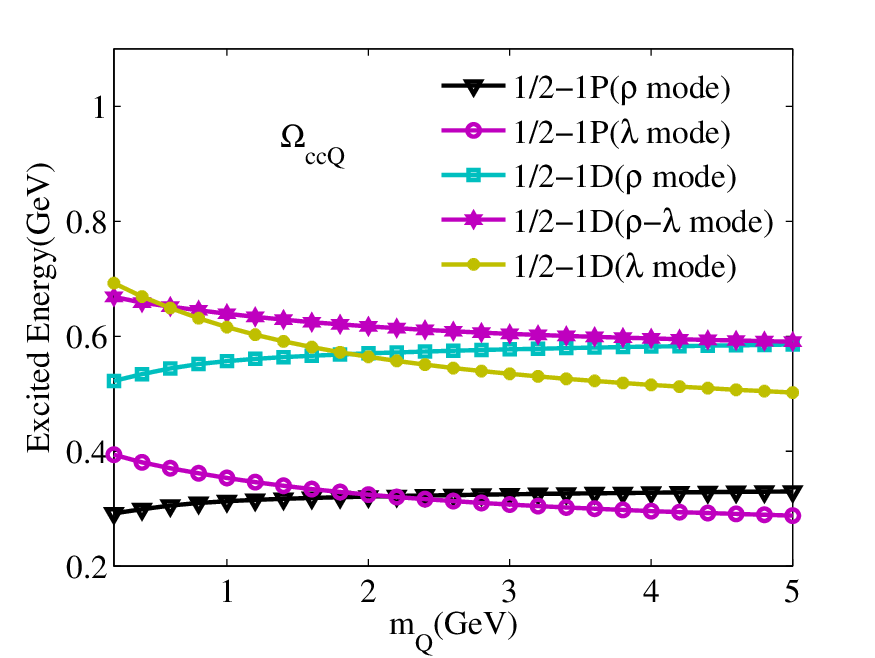}
  \end{minipage}
  }
 \subfigure[]{
   \begin{minipage}{6.5cm}
   \centering
   \includegraphics[width=6.5cm]{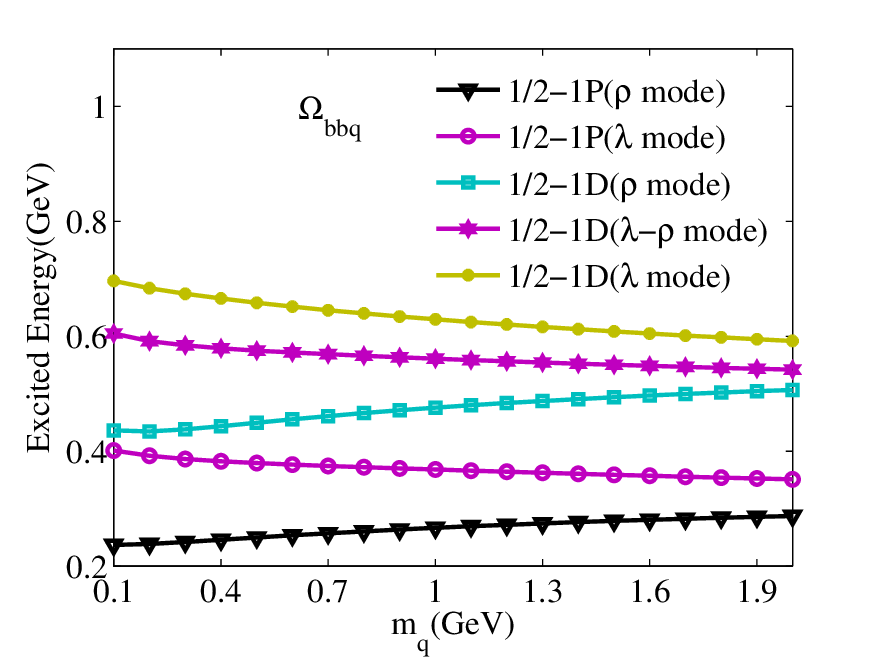}
  \end{minipage}
  }
  \caption{Quark mass dependence of the excited energy for 1P($\frac{1}{2}^{-}$) and 1D($\frac{1}{2}^{+}$) $\Omega_{ccQ}$ system (a) and $\Omega_{bbq}$ system (b).}
\label{mass}
\end{figure}
\begin{table}[h]
\begin{ruledtabular}\caption{Relevant parameters of the relativized quark model}
\label{parameter}
\begin{tabular}{c c c c c c}
$\sigma_{0}$(GeV)  &$\gamma_{1}$(GeV)&$\gamma_{2}$(GeV)&$\gamma_{3}$(GeV) &$b$(GeV$^{2}$)& $c$(MeV) \\
$1.8$ &$\frac{1}{2}$& $\sqrt{10}/2$ & $\sqrt{1000}/2$ & $0.14$ &$-198$ \\ \hline
$s$&$\epsilon_{\mathrm{c}}$ & $\epsilon_{\mathrm{(so)v}}$& $\epsilon_{\mathrm{t}}$ &$\epsilon_{\mathrm{(so)s}}$ &$\alpha_{1}$\\
$1.55$ & $-0.168$ & $-0.035$ &$0.025$ &$0.055$ &$0.25$\\ \hline
$\alpha_{2}$ &$\alpha_{3}$&$m_{c}$(MeV)&$m_{b}$(MeV)&  \\
$0.15$ &$0.20$&$1628$& $4997$ &    \\
\end{tabular}
\end{ruledtabular}
\end{table}
All of the interaction parameters in the Hamiltonian in Eq. (\ref{Hami}) are presented in Table \ref{parameter}. These parameters are taken as the same values as those in our previous works \cite{GLY1,ZYL1} where the experimental masses of singly heavy baryons were well reproduced. The orbital excitations of heavy baryons are usually classified into different modes according to the orbital angular momentum $l_{\rho}$ and $l_{\lambda}$. For $P$-wave baryons, they have two excitation modes which are called $\lambda$- and $\rho$-mode with ($l_{\rho}$,$l_{\lambda}$)=($0$,$1$) and ($1$,$0$), respectively. For $D$-wave baryons, there exist three types of excitation modes with ($l_{\rho}$,$l_{\lambda}$)=($0$,$2$), ($2$,$0$) and ($1$,$1$), which are called the $\lambda$-mode, $\rho$-mode and $\lambda$-$\rho$ mixing mode, respectively. For higher orbital excited states, their situations are similar to $D$-wave baryons which also have three excitation modes. By changing $m_{Q}$ from $0.1\sim5.0$ GeV for $\Omega_{ccQ}$ system, and $m_{q}$ from $0.1\sim2.0$ GeV for $\Omega_{bbq}$, we illustrate the quark mass dependence of excited energy for different excited modes in Fig. \ref{mass}. For $\Omega_{ccQ}$ system, it is explicitly shown that the $\lambda$-mode appears lower in excited energy than both the $\rho$-mode and $\lambda$-$\rho$ mixing mode with $m_{Q}\geq4$ GeV. This means that the lowest states of $\Omega_{ccb}$ baryons are dominated by the $\lambda$-mode. As for the $\Omega_{bbq}$ system, their excitations are dominated by $\rho$-mode, which are opposite to $\Omega_{ccQ}$ system. That is to say, the orbital excitation with the lowest energy is always associated with the heavier quark in the triply heavy baryons. This characteristic is consistent well with our previous conclusion which was named as the mechanism of heavy quark dominance \cite{Li:2023gbo}.

\begin{figure}[htbp]
  \centering
   \subfigure[]{
   \begin{minipage}{4cm}
   \centering
   \includegraphics[width=4.5cm]{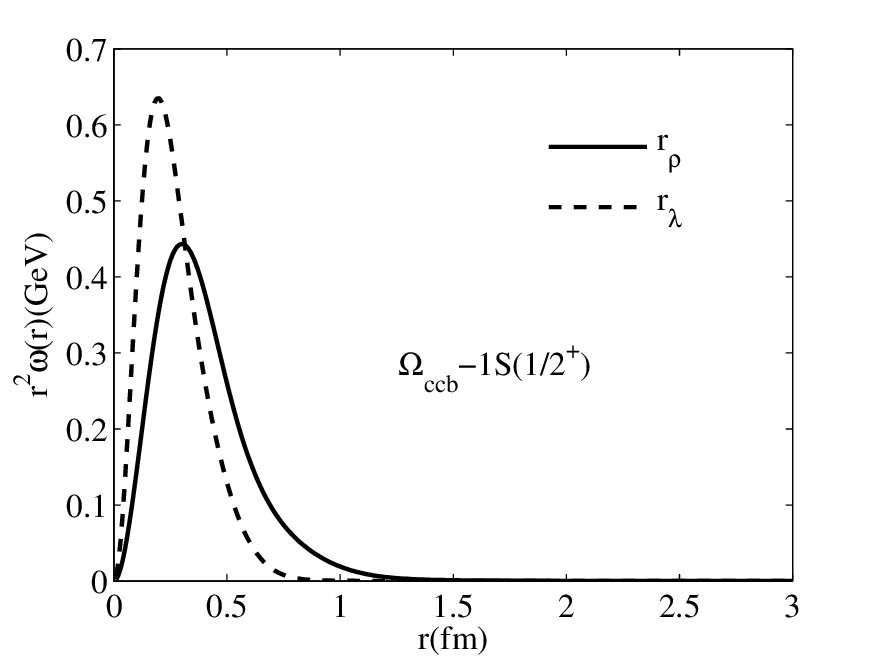}
  \end{minipage}
  }
 \subfigure[]{
   \begin{minipage}{4cm}
   \centering
   \includegraphics[width=4.5cm]{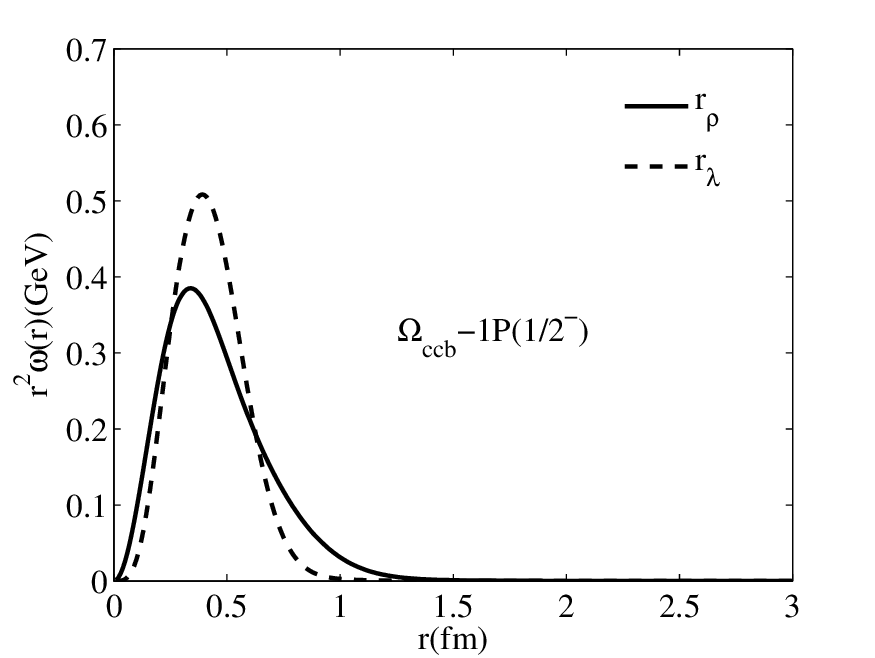}
  \end{minipage}
  }
 \subfigure[]{
   \begin{minipage}{4cm}
   \centering
   \includegraphics[width=4.5cm]{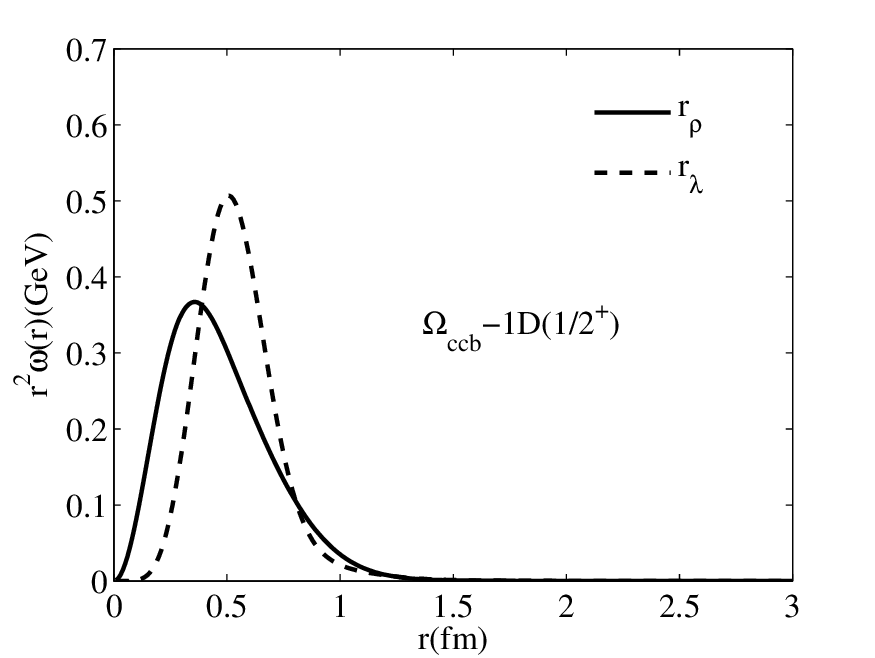}
  \end{minipage}
  }
  \subfigure[]{
   \begin{minipage}{4cm}
   \centering
   \includegraphics[width=4.5cm]{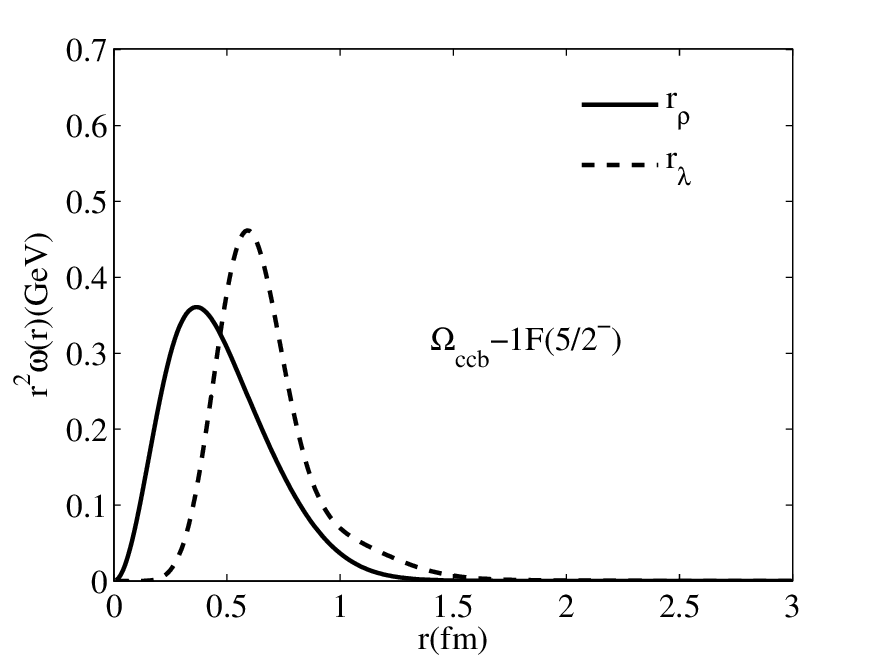}
  \end{minipage}
  }
  \caption{Radial density distributions for $1S\sim1F$ states in the $\Omega_{ccb}$ family with $\lambda-$mode.}
\label{ccborbital}
\end{figure}

\begin{figure}[htbp]
  \centering
   \subfigure[]{
   \begin{minipage}{4cm}
   \centering
   \includegraphics[width=4.5cm]{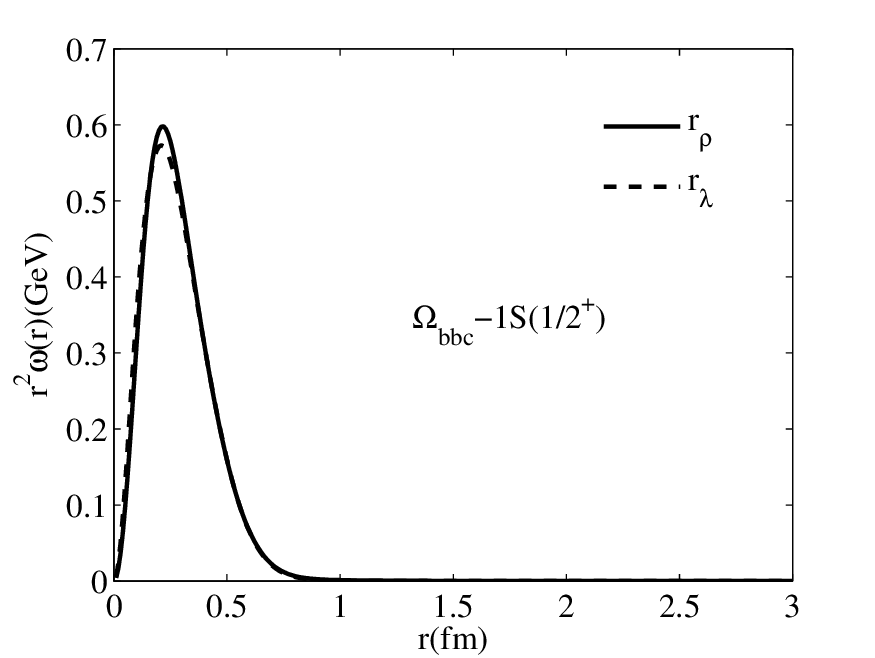}
  \end{minipage}
  }
 \subfigure[]{
   \begin{minipage}{4cm}
   \centering
   \includegraphics[width=4.5cm]{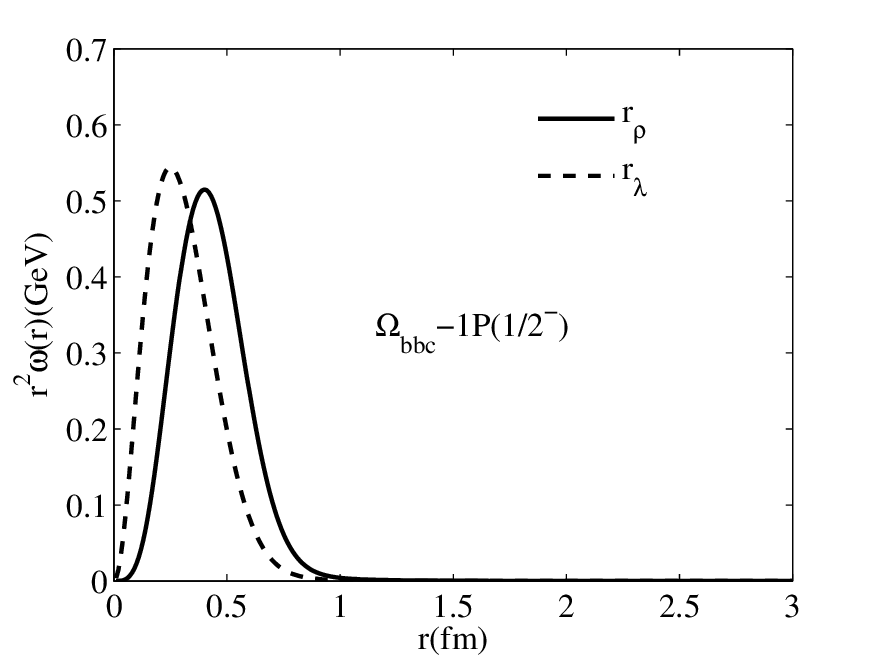}
  \end{minipage}
  }
 \subfigure[]{
   \begin{minipage}{4cm}
   \centering
   \includegraphics[width=4.5cm]{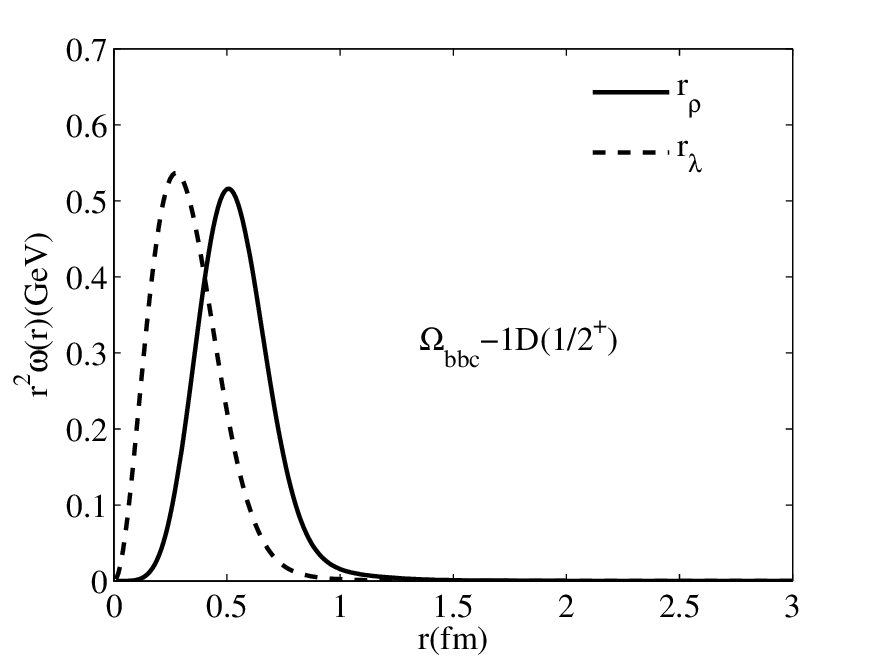}
  \end{minipage}
  }
  \subfigure[]{
   \begin{minipage}{4cm}
   \centering
   \includegraphics[width=4.5cm]{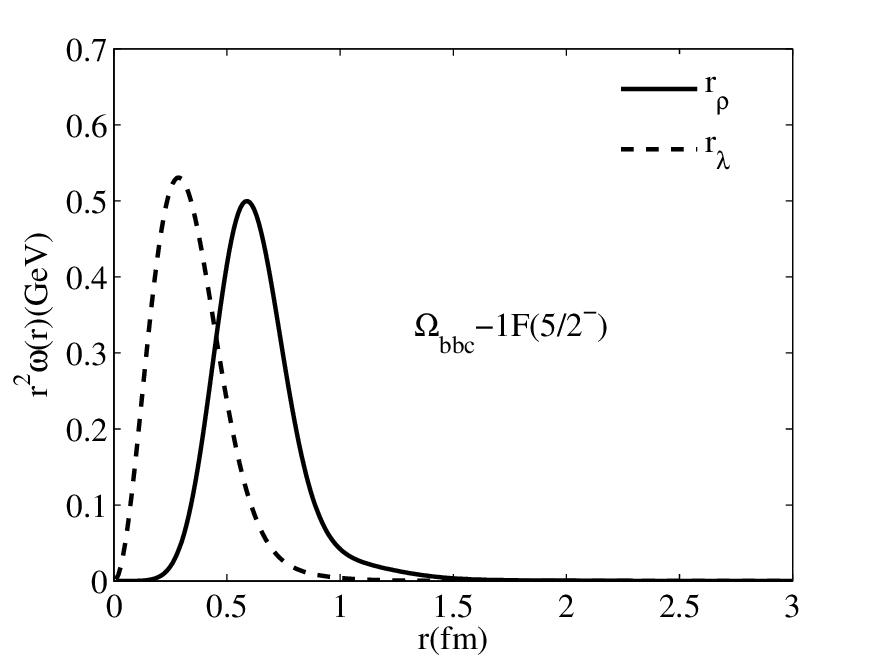}
  \end{minipage}
  }
  \caption{Radial density distributions for $1S\sim1F$ states in the $\Omega_{bbc}$ family with $\rho-$mode.}
\label{bbcorbital}
\end{figure}

\begin{figure}[htbp]
  \centering
   \subfigure[]{
   \begin{minipage}{4cm}
   \centering
   \includegraphics[width=4.5cm]{ccb1S.eps}
  \end{minipage}
  }
 \subfigure[]{
   \begin{minipage}{4cm}
   \centering
   \includegraphics[width=4.5cm]{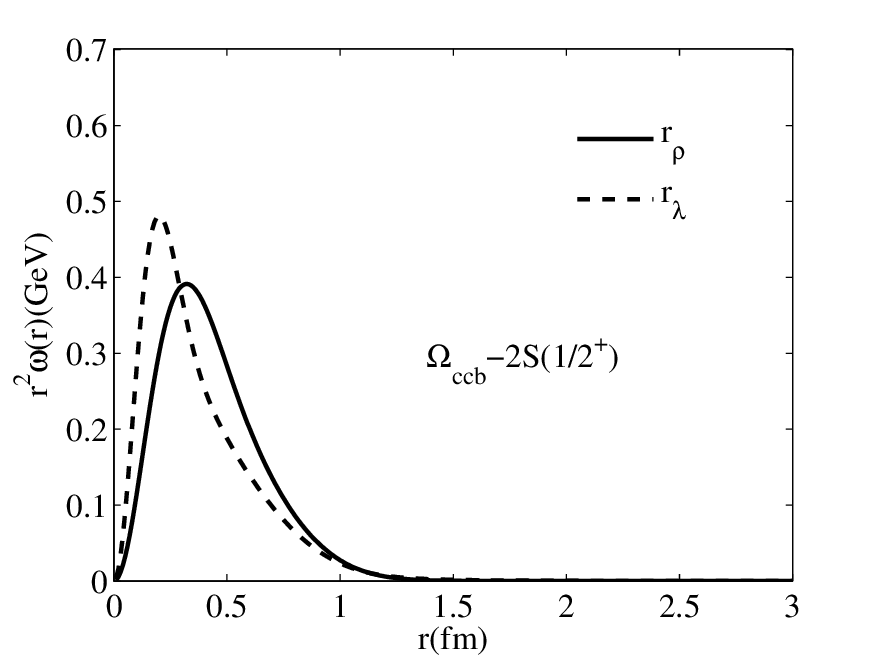}
  \end{minipage}
  }
 \subfigure[]{
   \begin{minipage}{4cm}
   \centering
   \includegraphics[width=4.5cm]{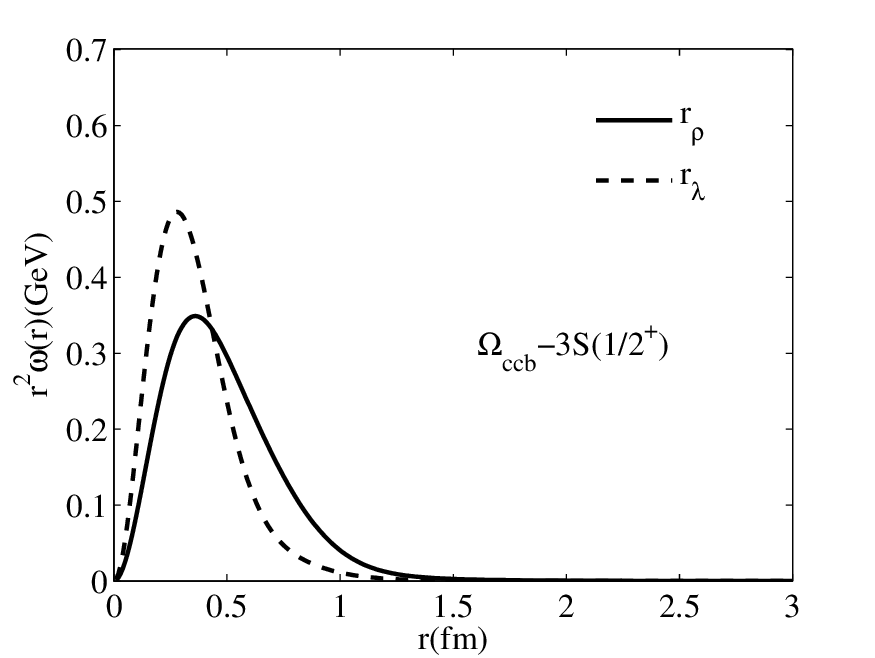}
  \end{minipage}
  }
  \caption{Radial density distributions for $1S\sim3S$ states in the $\Omega_{ccb}$ family.}
\label{ccbradial}
\end{figure}

For $\Omega_{ccb}$ with $\lambda$-mode and $\Omega_{bbc}$ with $\rho$-mode, we obtain their r.m.s. radii and mass spectra with quantum numbers up to $n=4$ and $L=4$. The results are listed in Tables \ref{ccbLambda} and \ref{bbcRho} in the Appendix. In order to further investigate the inner structure, we also analyze the radial density distribution of these triply heavy baryons. The radial density distributions are defined as,
\begin{eqnarray}
\notag
\omega(r_{\rho})=\int|\Psi(\textbf{r}_{\rho},\textbf{r}_{\lambda})|^{2}d\textbf{r}_{\lambda}d\Omega_{\rho} \\
\omega(r_{\lambda})=\int|\Psi(\textbf{r}_{\rho},\textbf{r}_{\lambda})|^{2}d\textbf{r}_{\rho}d\Omega_{\lambda}
\end{eqnarray}
where $\Omega_{\rho}$ and $\Omega_{\lambda}$ are the solid angles spanned by vectors $\textbf{r}_{\rho}$ and $\textbf{r}_{\lambda}$, respectively. Some of the results about the radial density distributions of baryons $\Omega_{ccb}$ and $\Omega_{bbc}$ are shown in Figs. \ref{ccborbital}-\ref{ccbradial}.

For $\Omega_{ccb}$ states with the same radial quantum number $n$, their $\sqrt{\langle r_{\lambda}^{2}\rangle}$ becomes larger obviously when the orbital angular momentum $L$ increases (see Table \ref{ccbLambda}). However, $ \sqrt{\langle r_{\rho}^{2}\rangle}$ increases a little with $L$ increasing. The situation is opposite to $\Omega_{bbc}$ states whose values of $ \sqrt{\langle r_{\rho}^{2}\rangle}$ increase more quickly with the orbital angular $L$ than those of $\sqrt{\langle r_{\lambda}^{2}\rangle}$ (see Table \ref{bbcRho}). Figs. \ref{ccborbital}-\ref{bbcorbital} also show similar characteristic about the radial density distribution. It is shown that the $r^{2}\omega(r_{\lambda})$ peak of $\Omega_{ccb}$ states shifts outward more evidently than that of $r^{2}\omega(r_{\rho})$ with $L$ increment. However, the situation is opposite to $\Omega_{bbc}$ baryons. These above phenomenons can be well explained by $\Omega_{ccb}$ and $\Omega_{bbc}$ baryons having different orbital excited modes. Because dominant orbital excitations is $\lambda-$mode for $\Omega_{ccb}$ baryon, this makes its $\sqrt{\langle r_{\lambda}^{2}\rangle}$ increase faster and $r^{2}\omega(r_{\lambda})$ peak shift outward more quickly. As for $\Omega_{bbc}$ system, its situation is exactly opposite to the former. For these states with the same angular momentum $L$, Tables \ref{ccbLambda} and \ref{bbcRho} show that both $\sqrt{\langle r_{\rho}^{2}\rangle}$ and $\sqrt{\langle r_{\lambda}^{2}\rangle}$ increase with radial quantum number $n$. We can also see this feature from Fig. \ref{ccbradial}, where the peak of radial density distribution becomes lower from $1S\sim3S$ states and the peak position shifts outward slightly. Theoretically, the larger the r.m.s. radii become, the looser the baryons will be. We hope these results can help to estimate the upper limit of the mass spectra and to search for the $\Omega_{ccb}$ and $\Omega_{bbc}$ baryons in forthcoming experiments.

\subsection{ Mass spectra of $\Omega_{ccb}$ and $\Omega_{bbc}$ baryons}

\begin{table}[htbp]
\begin{ruledtabular}\caption{Predicted masses(in MeV) of the $1P(\frac{3}{2}^{-})$ and $1D(\frac{5}{2}^{+})$ $\Omega_{ccb}$ heavy baryon.}
\label{mixtrue}
\begin{tabular}{c c c| c c }
\multicolumn{3}{c|}{Single configuration}& \multicolumn{2}{c}{Configuration mixing} \\ \hline
$nL$($J^{P}$) & $l_{\rho}$ $l_{\lambda}$ L s j & Mass &  Eigenvalues  &  Mixing coefficients($\%$)   \\  \hline
\multirow{3}{*}{$1P(\frac{3}{2}^{-})$}
               & 0 1 1 1 1 & 8319 & 8302 & (\textbf{34.9}, \textbf{64.1}, 1.0)  \\
              ~& 0 1 1 1 2 & 8311 & 8327 & (\textbf{65.0}, \textbf{33.8}, 1.2) \\
              ~& 1 0 1 0 1 & 8370 & 8370 & (1.1, 0.8, \textbf{98.1})   \\  \hline
\multirow{5}{*}{$1D(\frac{5}{2}^{+})$}
               & 0 2 2 1 2 & 8532 & 8518 & (\textbf{39.7}, \textbf{60.0}, 0.1, 0.1, 0.1)  \\
               & 0 2 2 1 3 & 8527 & 8541 & (\textbf{59.8}, \textbf{39.9}, 0.1, 0.1, 0.1)  \\
               & 1 1 2 0 2 & 8585 & 8585 & (0.5, 0.5, \textbf{98.6}, 0.2, 0.2)  \\
              ~& 2 0 2 1 2 & 8629 & 8615 & (0.1, 0.1, 0.4, 0.4, \textbf{99.0}) \\
              ~& 2 0 2 1 3 & 8615 & 8629 & (0.1, 0.1, 0.3, \textbf{99.0}, 0.6)   \\
\end{tabular}
\end{ruledtabular}
\end{table}
\begin{table*}[htbp]
\begin{ruledtabular}\caption{Predicted masses(in MeV) of the $\Omega_{ccb}$ baryons.}
\label{massccb}
\begin{tabular}{c| c c c c c c c c c}
&$nL(J^{P})$&This work &\cite{Yang:2019lsg} &\cite{Silves:1996myf} & \cite{Serafin:2018aih} &\cite{Wang:2011ae,Mathur:2018epb}&\cite{Qin:2019hgk}&\cite{Flynn:2011gf}&\cite{Flynn:2011gf}\\ \hline
\multirow{8}{*}{S-wave}&$1S$($\frac{1}{2}^{+}$) &  8025 & 8004 & 8019 & 8301 & 8005(13)&7867&8018 & 8058\\
&$2S$($\frac{1}{2}^{+}$) &  8422 & 8455 & 8450 & 8600 & &8337  \\
&$3S$($\frac{1}{2}^{+}$) &  8522 &      &      &      & &      \\
&$4S$($\frac{1}{2}^{+}$) &  8731 &      &      &      & &      \\ \cline{2-10}	
&$1S$($\frac{3}{2}^{+}$) &  8046 & 8023 & 8056 & 8301 &8026(13)& 7963 & 8046&8087\\
&$2S$($\frac{3}{2}^{+}$) &  8438 & 8468 & 8465 & 8600 & &8427  \\
&$3S$($\frac{3}{2}^{+}$) &  8563 &      &      &      &   &    \\
&$4S$($\frac{3}{2}^{+}$) &  8745 &      &      &      \\ \hline
\multirow{12}{*}{P-wave}&$1P$($\frac{1}{2}^{-}$) &  8303 & 8306 & 8316 & 8491 & 8360(130) & 8164 \\
&$2P$($\frac{1}{2}^{-}$) &  8611 & 8663 & 8579 &  \\
&$3P$($\frac{1}{2}^{-}$) &  8738 &      &      &      \\
&$4P$($\frac{1}{2}^{-}$) &  8881 &      &      &      \\ \cline{2-10}
&$1P$($\frac{3}{2}^{-}$) &  8302 & 8306 & 8316 & 8491 & 8360(130)&8275 \\
&$2P$($\frac{3}{2}^{-}$) &  8609 & 8663 & 8579 &  \\
&$3P$($\frac{3}{2}^{-}$) &  8738 &      &      &      \\
&$4P$($\frac{3}{2}^{-}$) &  8878 &      &      &      \\ \cline{2-10}
&$1P$($\frac{5}{2}^{-}$) &  8321 & 8311 & 8331 & 8491 \\
&$2P$($\frac{5}{2}^{-}$) &  8637 & 8667 & 8589 &  \\
&$3P$($\frac{5}{2}^{-}$) &  8749 &      &      &      \\
&$4P$($\frac{5}{2}^{-}$) &  8919 &      &      &      \\ \hline
\multirow{16}{*}{D-wave}&$1D$($\frac{1}{2}^{+}$) &  8524 & 8536 & 8528 & 8647 \\
&$2D$($\frac{1}{2}^{+}$) &  8798 & 8838 & 8762 &  \\
&$3D$($\frac{1}{2}^{+}$) &  8914 &      &      &      \\
&$4D$($\frac{1}{2}^{+}$) &  9076 &      &      &      \\ \cline{2-10}
&$1D$($\frac{3}{2}^{+}$) &  8525 & 8536 & 8528 & 8647 \\
&$2D$($\frac{3}{2}^{+}$) &  8788 & 8838 & 8762 &  \\
&$3D$($\frac{3}{2}^{+}$) &  8914 &      &      &      \\
&$4D$($\frac{3}{2}^{+}$) &  9045 &      &      &      \\ \cline{2-10}
&$1D$($\frac{5}{2}^{+}$) &  8518 & 8536 & 8528 & 8647 \\
&$2D$($\frac{5}{2}^{+}$) &  8758 & 8838 & 8762 &  \\
&$3D$($\frac{5}{2}^{+}$) &  8912 &      &      &      \\
&$4D$($\frac{5}{2}^{+}$) &  9020 &      &      &      \\ \cline{2-10}
&$1D$($\frac{7}{2}^{+}$) &  8532 & 8538 & 8528 & 8647 \\
&$2D$($\frac{7}{2}^{+}$) &  8802 & 8839 & 8762 &  \\
&$3D$($\frac{7}{2}^{+}$) &  8918 &      &      &      \\
&$4D$($\frac{7}{2}^{+}$) &  9106 &      &      &      \\ \hline
\multirow{16}{*}{F-wave}&$1F$($\frac{1}{2}^{-}$) &  8748 &     &     &     \\
&$2F$($\frac{1}{2}^{-}$) &  9009 &      &      &      \\
&$3F$($\frac{1}{2}^{-}$) &  9089 &      &      &      \\
&$4F$($\frac{1}{2}^{-}$) &  9270 &      &      &      \\ \cline{2-10}
&$1F$($\frac{3}{2}^{-}$) &  8707 &      &      &      \\
&$2F$($\frac{3}{2}^{-}$) &  8941 &      &      &      \\
&$3F$($\frac{3}{2}^{-}$) &  9071 &      &      &      \\
&$4F$($\frac{3}{2}^{-}$) &  9272 &      &      &      \\ \cline{2-10}
&$1F$($\frac{5}{2}^{-}$) &  8705 &      &      &      \\
&$2F$($\frac{5}{2}^{-}$) &  8902 &      &      &      \\
&$3F$($\frac{5}{2}^{-}$) &  9070 &      &      &      \\
&$4F$($\frac{5}{2}^{-}$) &  9267 &      &      &      \\ \cline{2-10}
&$1F$($\frac{7}{2}^{-}$) &  8704 &      &      &      \\
&$2F$($\frac{7}{2}^{-}$) &  8899 &      &      &      \\
&$3F$($\frac{7}{2}^{-}$) &  9070 &      &      &      \\
&$4F$($\frac{7}{2}^{-}$) &  9270 &      &      &      \\
\end{tabular}
\end{ruledtabular}
\end{table*}
\begin{table*}[htbp]
\begin{ruledtabular}\caption{Predicted masses(in MeV) of the $\Omega_{bbc}$ baryons.}
\label{massbbc}
\begin{tabular}{c| c c c c c c c c c}
			$l_{\rho}$  $l_{\lambda}$ L s j & $nL$($J^{P}$) &This work &\cite{Yang:2019lsg} &\cite{Silves:1996myf} & \cite{Serafin:2018aih}  &\cite{Mathur:2018epb}&\cite{Qin:2019hgk}&\cite{Flynn:2011gf}&\cite{Flynn:2011gf}\\ \hline
			\multirow{8}{*}{S-wave }
			~& $1S$($\frac{1}{2}^{+}$)  & 11217 & 11200 & 11217 & 11218 & 11500(110) &11077&11214 & 11247\\
			~& $2S$($\frac{1}{2}^{+}$)  & 11604 & 11607 & 11625 & 11585 & & 11603\\
			~& $3S$($\frac{1}{2}^{+}$)  & 11700  &  &  & \\
			~& $4S$($\frac{1}{2}^{+}$)  & 11888 &  &  & \\ \cline{2-10}
			
			~ & $1S$($\frac{3}{2}^{+}$) & 11236 & 11221 & 11251 & 11218 &11490(110)&11167&11245 & 11281\\
			~ & $2S$($\frac{3}{2}^{+}$) & 11617 & 11622 & 11643 & 11585 & &11703\\
			~ & $3S$($\frac{3}{2}^{+}$) & 11709 &  &  &  \\
			~ & $4S$($\frac{3}{2}^{+}$) & 11899 &  &  & \\ \hline
			\multirow{12}{*}{P-wave}
			~ & $1P$($\frac{1}{2}^{-}$)  & 11492 & 11482 & 11524 & 11438 &11620(110) &11413\\
			~ & $2P$($\frac{1}{2}^{-}$)  & 11798 & 11802 & 11820 & \\
			~ & $3P$($\frac{1}{2}^{-}$)  & 11900 &  &  & \\
			~ & $4P$($\frac{1}{2}^{-}$)  & 12046 &  &  &  \\ \cline{2-10}
			
			~ & $1P$($\frac{3}{2}^{-}$)  & 11506 & 11482 & 11524 & 11438 &11620(110)&11523\\
			~ & $2P$($\frac{3}{2}^{-}$)  & 11809 & 11802 & 11820 & \\
			~ & $3P$($\frac{3}{2}^{-}$)  & 11900 &  &  & \\
			~ & $4P$($\frac{3}{2}^{-}$)  & 12057 &  &  & \\ \cline{2-10}

			~ & $1P$($\frac{5}{2}^{-}$)  & 11562 & 11569 & 11598 & 11601 \\
			~ & $2P$($\frac{5}{2}^{-}$)  & 11881 & 11888 & 11899 & \\
			~ & $3P$($\frac{5}{2}^{-}$)  & 11909 &  &  & \\
			~ & $4P$($\frac{5}{2}^{-}$)  & 12138 &  &  & \\ \hline
			\multirow{16}{*}{D-wave}
			~ & $1D$($\frac{1}{2}^{+}$)  & 11690 & 11677 & 11718 & 11626 \\
			~ & $2D$($\frac{1}{2}^{+}$)  & 11960 & 11955 & 11986 &  \\
			~ & $3D$($\frac{1}{2}^{+}$)  & 12090 &  &  & \\
			~ & $4D$($\frac{1}{2}^{+}$)  & 12209 &  &  & \\  \cline{2-10}
			
			~ & $1D$($\frac{3}{2}^{+}$)  & 11688 & 11677 & 11718 & 11626 \\
			~ & $2D$($\frac{3}{2}^{+}$)  & 11959 & 11955 & 11986 &    \\
			~ & $3D$($\frac{3}{2}^{+}$)  & 12100 &  &  & \\
			~ & $4D$($\frac{3}{2}^{+}$)  & 12208 &  &  & \\ \cline{2-10}
			
			~ & $1D$($\frac{5}{2}^{+}$)  & 11688 & 11677 & 11718 & 11626 \\
			~ & $2D$($\frac{5}{2}^{+}$)  & 11959 & 11955 & 11986 & \\
			~ & $3D$($\frac{5}{2}^{+}$)  & 12100 &  &  & \\
			~ & $4D$($\frac{5}{2}^{+}$)  & 12211 &  &  & \\ \cline{2-10}
			
			~ & $1D$($\frac{7}{2}^{+}$)  & 11713 & 11688 & 11718 & 11626 \\
			~ & $2D$($\frac{7}{2}^{+}$)  & 11979 & 11963 & 11986 &  \\
			~ & $3D$($\frac{7}{2}^{+}$)  & 12123 &  &  & \\
			~ & $4D$($\frac{7}{2}^{+}$)  & 12237 &  &  & \\ \hline
			\multirow{16}{*}{F-wave}
		    ~ & $1F$($\frac{1}{2}^{-}$)  & 11920 &  &  &  \\
			~ & $2F$($\frac{1}{2}^{-}$)  & 12146 &  &  &  \\
			~ & $3F$($\frac{1}{2}^{-}$)  & 12259 &  &  & \\
			~ & $4F$($\frac{1}{2}^{-}$)  & 12420 &  &  & \\  \cline{2-10}

		    ~ & $1F$($\frac{3}{2}^{-}$)  & 11921 &  &  &  \\
			~ & $2F$($\frac{3}{2}^{-}$)  & 12147 &  &  &  \\
			~ & $3F$($\frac{3}{2}^{-}$)  & 12260 &  &  & \\
			~ & $4F$($\frac{3}{2}^{-}$)  & 12422 &  &  & \\  \cline{2-10}

			~ & $1F$($\frac{5}{2}^{-}$)  & 11854 &  &  &  \\
			~ & $2F$($\frac{5}{2}^{-}$)  & 12097 &  &  &  \\
			~ & $3F$($\frac{5}{2}^{-}$)  & 12250 &  &  & \\
			~ & $4F$($\frac{5}{2}^{-}$)  & 12380 &  &  & \\  \cline{2-10}
			
			~ & $1F$($\frac{7}{2}^{-}$)  & 11875 &  &  &  \\
			~ & $2F$($\frac{7}{2}^{-}$)  & 12114 &  &  &  \\
			~ & $3F$($\frac{7}{2}^{-}$)  & 12265 &  &  & \\
			~ & $4F$($\frac{7}{2}^{-}$)  & 12403 &  &  &
\end{tabular}
\end{ruledtabular}
\end{table*}
\begin{table*}[htbp]
\begin{ruledtabular}\caption{Predicted masses(in MeV) of the $\Omega_{ccc}$ baryons.}
\label{massccc}
\begin{tabular}{c| c c c c c c c c c}
&$nL(J^{P})$&This work &\cite{Yang:2019lsg} &\cite{Silves:1996myf} & \cite{Serafin:2018aih} &\cite{Mathur:2018epb}&\cite{Qin:2019hgk}&\cite{Flynn:2011gf}&\cite{Flynn:2011gf}\\ \hline
\multirow{4}{*}{S-wave}&$1S$($\frac{3}{2}^{+}$) &  4805 & 4798 & 4799 & 4797 & 4759(6)&4760&4799&4847\\
&$2S$($\frac{3}{2}^{+}$) &  5219 & 5286 & 5243 & 5309 & 5313(31)&5150  \\
&$3S$($\frac{3}{2}^{+}$) &  5317 &      &      &      & &      \\
&$4S$($\frac{3}{2}^{+}$) &  5569 &      &      &      & &      \\ \hline	
\multirow{12}{*}{P-wave}&$1P$($\frac{1}{2}^{-}$) &  5083 & 5129 & 5094 & 5103 & 5116(9) & \\
&$2P$($\frac{1}{2}^{-}$) &  5425 & 5525 & 5456 &  & 5608(31)\\
&$3P$($\frac{1}{2}^{-}$) &  5515 &      &      &      \\
&$4P$($\frac{1}{2}^{-}$) &  5745 &      &      &      \\ \cline{2-10}
&$1P$($\frac{3}{2}^{-}$) &  5091 & 5129 & 5094 & 5103 & 5120(13)&5027 \\
&$2P$($\frac{3}{2}^{-}$) &  5426 & 5525 & 5456 &  &5658(31)\\
&$3P$($\frac{3}{2}^{-}$) &  5514 &      &      &      \\
&$4P$($\frac{3}{2}^{-}$) &  5750 &      &      &      \\ \cline{2-10}
&$1P$($\frac{5}{2}^{-}$) &  5114 & 5558 & 5494 &    &5512(64)  \\
&$2P$($\frac{5}{2}^{-}$) &  5453 & 5846 &      &  &5705(25)\\
&$3P$($\frac{5}{2}^{-}$) &  5529 &      &      &      \\
&$4P$($\frac{5}{2}^{-}$) &  5775 &      &      &      \\ \hline
\multirow{16}{*}{D-wave}&$1D$($\frac{1}{2}^{+}$) &  5313 & 5376 & 5324 & 5358 &  5395(13)\\
&$2D$($\frac{1}{2}^{+}$) &  5620 & 5713 &      &  \\
&$3D$($\frac{1}{2}^{+}$) &  5706 &      &      &      \\
&$4D$($\frac{1}{2}^{+}$) &  5887 &      &      &      \\ \cline{2-10}
&$1D$($\frac{3}{2}^{+}$) &  5330 & 5376 & 5324 & 5358 & 5426(13)\\
&$2D$($\frac{3}{2}^{+}$) &  5629 & 5713 &      &  \\
&$3D$($\frac{3}{2}^{+}$) &  5723 &      &      &      \\
&$4D$($\frac{3}{2}^{+}$) &  5911 &      &      &      \\ \cline{2-10}
&$1D$($\frac{5}{2}^{+}$) &  5329 & 5376 & 5324 & 5358 & 5402(15)\\
&$2D$($\frac{5}{2}^{+}$) &  5602 & 5713 &      &  \\
&$3D$($\frac{5}{2}^{+}$) &  5721 &      &      &      \\
&$4D$($\frac{5}{2}^{+}$) &  5917 &      &      &      \\ \cline{2-10}
&$1D$($\frac{7}{2}^{+}$) &  5353 & 5376 & 5324 & 5358 & 5393(49)\\
&$2D$($\frac{7}{2}^{+}$) &  5648 & 5713 &      &  \\
&$3D$($\frac{7}{2}^{+}$) &  5727 &      &      &      \\
&$4D$($\frac{7}{2}^{+}$) &  5947 &      &      &      \\ \hline
\multirow{16}{*}{F-wave}&$1F$($\frac{1}{2}^{-}$) &  5545 &     &     &     \\
&$2F$($\frac{1}{2}^{-}$) &  5837 &      &      &      \\
&$3F$($\frac{1}{2}^{-}$) &  5899 &      &      &      \\
&$4F$($\frac{1}{2}^{-}$) &  6079 &      &      &      \\ \cline{2-10}
&$1F$($\frac{3}{2}^{-}$) &  5548 &      &      &      \\
&$2F$($\frac{3}{2}^{-}$) &  5825 &      &      &      \\
&$3F$($\frac{3}{2}^{-}$) &  5902 &      &      &      \\
&$4F$($\frac{3}{2}^{-}$) &  6082 &      &      &      \\ \cline{2-10}
&$1F$($\frac{5}{2}^{-}$) &  5534 &      &      &      \\
&$2F$($\frac{5}{2}^{-}$) &  5738 &      &      &      \\
&$3F$($\frac{5}{2}^{-}$) &  5902 &      &      &      \\
&$4F$($\frac{5}{2}^{-}$) &  6079 &      &      &      \\ \cline{2-10}
&$1F$($\frac{7}{2}^{-}$) &  5535 &      &      &      \\
&$2F$($\frac{7}{2}^{-}$) &  5758 &      &      &      \\
&$3F$($\frac{7}{2}^{-}$) &  5902 &      &      &      \\
&$4F$($\frac{7}{2}^{-}$) &  6083 &      &      &      \\
\end{tabular}
\end{ruledtabular}
\end{table*}
\begin{table*}[htbp]
\begin{ruledtabular}\caption{Predicted masses(in MeV) of the $\Omega_{bbb}$ baryons.}
\label{massbbb}
\begin{tabular}{c| c c c c c c c c c}
&$nL(J^{P})$&This work &\cite{Yang:2019lsg} &\cite{Silves:1996myf} & \cite{Serafin:2018aih} &\cite{Mathur:2018epb}&\cite{Qin:2019hgk}&\cite{Flynn:2011gf}&\cite{Flynn:2011gf}\\ \hline
\multirow{4}{*}{S-wave}&$1S$($\frac{3}{2}^{+}$) &  14394 & 14396 & 14398 & 14347 & 14371(12)&14370& 14398 & 14424 \\
&$2S$($\frac{3}{2}^{+}$) &  14782 & 14805 & 14835 & 14832 & 14840(14) & 14980 \\
&$3S$($\frac{3}{2}^{+}$) &  14873 &      &      &      & &      \\
&$4S$($\frac{3}{2}^{+}$) &  15079 &      &      &      & &      \\ \hline	
\multirow{12}{*}{P-wave}&$1P$($\frac{1}{2}^{-}$) &  14682 & 14688 & 14738 & 14645 & 14706(9) & 8164 \\
&$2P$($\frac{1}{2}^{-}$) &  14984 & 15016 & 15052 &  \\
&$3P$($\frac{1}{2}^{-}$) &  15053 &      &      &      \\
&$4P$($\frac{1}{2}^{-}$) &  15218 &      &      &      \\ \cline{2-10}
&$1P$($\frac{3}{2}^{-}$) &  14683 & 14688 & 14738 & 14645 & 14714(9)&14771 \\
&$2P$($\frac{3}{2}^{-}$) &  14982 & 15016 & 15052 &  \\
&$3P$($\frac{3}{2}^{-}$) &  15052 &      &      &      \\
&$4P$($\frac{3}{2}^{-}$) &  15217 &      &      &      \\ \cline{2-10}
&$1P$($\frac{5}{2}^{-}$) &  14693 & 15038 & 15078 &   \\
&$2P$($\frac{5}{2}^{-}$) &  14992 & 15284 & 15402 &  \\
&$3P$($\frac{5}{2}^{-}$) &  15058 &      &      &      \\
&$4P$($\frac{5}{2}^{-}$) &  15233 &      &      &      \\ \hline
\multirow{16}{*}{D-wave}&$1D$($\frac{1}{2}^{+}$) &  14873 & 14894 & 14944 & 14896 & 14938(18)\\
&$2D$($\frac{1}{2}^{+}$) &  15138 & 15175 & 15304 &  \\
&$3D$($\frac{1}{2}^{+}$) &  15215 &      &      &      \\
&$4D$($\frac{1}{2}^{+}$) &  15357 &      &      &      \\ \cline{2-10}
&$1D$($\frac{3}{2}^{+}$) &  14900 & 14894 & 14944 & 14896 &14958(18)\\
&$2D$($\frac{3}{2}^{+}$) &  15147 & 15175 & 15304 &  \\
&$3D$($\frac{3}{2}^{+}$) &  15223 &      &      &      \\
&$4D$($\frac{3}{2}^{+}$) &  15332 &      &      &      \\ \cline{2-10}
&$1D$($\frac{5}{2}^{+}$) &  14896 & 14894 & 14944 & 14896 & 14964(18)\\
&$2D$($\frac{5}{2}^{+}$) &  15135 & 15175 & 15304 &  \\
&$3D$($\frac{5}{2}^{+}$) &  15222 &      &      &      \\
&$4D$($\frac{5}{2}^{+}$) &  15297 &      &      &      \\ \cline{2-10}
&$1D$($\frac{7}{2}^{+}$) &  14904 & 14894 & 14944 & 14896 & 14969(17)\\
&$2D$($\frac{7}{2}^{+}$) &  15163 & 15175 & 15304 &  \\
&$3D$($\frac{7}{2}^{+}$) &  15225 &      &      &      \\
&$4D$($\frac{7}{2}^{+}$) &  15359 &      &      &      \\ \hline
\multirow{16}{*}{F-wave}&$1F$($\frac{1}{2}^{-}$) &  15075 &     &     &     \\
&$2F$($\frac{1}{2}^{-}$) &  15317 &      &      &      \\
&$3F$($\frac{1}{2}^{-}$) &  15375 &      &      &      \\
&$4F$($\frac{1}{2}^{-}$) &  15544 &      &      &      \\ \cline{2-10}
&$1F$($\frac{3}{2}^{-}$) &  15069 &      &      &      \\
&$2F$($\frac{3}{2}^{-}$) &  15313 &      &      &      \\
&$3F$($\frac{3}{2}^{-}$) &  15371 &      &      &      \\
&$4F$($\frac{3}{2}^{-}$) &  15486 &      &      &      \\ \cline{2-10}
&$1F$($\frac{5}{2}^{-}$) &  15068 &      &      &      \\
&$2F$($\frac{5}{2}^{-}$) &  15300 &      &      &      \\
&$3F$($\frac{5}{2}^{-}$) &  15371 &      &      &      \\
&$4F$($\frac{5}{2}^{-}$) &  15486 &      &      &      \\  \cline{2-10}
&$1F$($\frac{7}{2}^{-}$) &  15067 &      &      &      \\
&$2F$($\frac{7}{2}^{-}$) &  15304 &      &      &      \\
&$3F$($\frac{7}{2}^{-}$) &  15371 &      &      &      \\
&$4F$($\frac{7}{2}^{-}$) &  15487 &      &      &      \\
\end{tabular}
\end{ruledtabular}
\end{table*}

Based on the mechanism of heavy quark dominance, the energies of $\Omega_{ccb}$ baryons with $\lambda-$mode and $\Omega_{bbc}$ with $\rho-$mode are good approximations to their mass spectra. However, all possible assignments of the angular momenta with the same quantum number $J^{P}$ should also contribute to the mass spectra of the triply baryons. For $\Omega_{ccb}$ as an example, all of the possible assignments for 1$P(\frac{3}{2}^{-})$ and 1$D(\frac{5}{2}^{+})$ are listed in Table \ref{mixtrue}. From this table, we can see that the energies of the single configuration with $\lambda-$mode are truly lower than the other configurations. For example, the configurations ($l_{\rho}$ $l_{\lambda}$ $L$ $s$ $j$)=($0$ $1$ $1$ $1$ $1$), ($0$ $1$ $1$ $1$ $2$) for 1$P(\frac{3}{2}^{-})$ state and ($0$ $2$ $2$ $1$ $2$), ($0$ $2$ $2$ $1$ $3$) for 1$D(\frac{5}{2}^{+})$ state are lower states in energy than the others. We also calculate the eigenvalues and mixing coefficients by considering the configurations mixing. The results are shown in the last two columns in Table \ref{mixtrue}. It is shown that the lowest energy for 1$P(\frac{3}{2}^{-})$ state is $8311$ MeV without considering the mixing effect. This value becomes to be $8302$ MeV after considering the configuration mixing. For 1$D(\frac{5}{2}^{+})$ state, this value changes from $8527$ MeV to $8518$ MeV. That is to say, the lowest energy for each $J^{P}$ state is slightly lowered if the configuration mixing is considered.

\begin{table*}[htbp]
\begin{ruledtabular}\caption{Predicted masses(in MeV) of the $1P(\frac{5}{2}^{-})$ and $1F(\frac{3}{2}^{-})$.}
\label{ccb-bbc}
\begin{tabular}{c c| c| c c| c| c c}
\multicolumn{2}{c|}{Configuration}& \multicolumn{3}{c|}{$\Omega_{ccb}$} & \multicolumn{3}{c}{$\Omega_{bbc}$} \\ \hline
$nL$($J^{P}$) & $l_{\rho}$ $l_{\lambda}$ L s j & Mass &  Eigenvalues  &  Mixing coefficients($\%$) & Mass &  Eigenvalues  &  Mixing coefficients($\%$)  \\  \hline
$1P(\frac{5}{2}^{-})$ & 0 1 1 1 2 & 8321 & 8.321 & (100) & 11562 & 11562 & (100) \\ \hline
\multirow{8}{*}{$1F(\frac{3}{2}^{-})$}
               & 0 3 3 1 2 & 8707 & 8707 & (\textbf{99.9}, 0.1, 0.0, 0.0, 0.0, 0.0, 0.0, 0.0) & 11992 & 11921 & (0.0, 0.0, 0.0, \textbf{88.9}, \textbf{10.6}, 0.1, 4.9, 0.0) \\
              ~& 1 2 1 0 1 & 8752 & 8752 & (0.1, \textbf{99.8}, 0.1, 0.0, 0.0, 0.0, 0.0, 0.0) & 11993 & 11926 & (0.0, 0.0, 0.0, \textbf{10.8}, \textbf{89.1}, 0.0, 0.1, 0.0) \\
              ~& 1 2 2 0 2 & 8801 & 8773 & (0.0, 0.0, 0.0,\textbf{98.2}, 0.2 ,0.0, 0.6, 0.0)  & 12020 & 11936 &  (0.0, 0.0, 0.0, 0.0, 0.0, 0.4, 0.2, \textbf{99.4}) \\
               & 2 1 1 1 1 & 8773 & 8779 & (0.0, 0.0, 0.0, 1.5,\textbf{98.2}, 0.0, 0.3, 0.0) & 11922 & 11973 &  (0.0, 0.0, 0.0, 0.3, 0.3, \textbf{14.7}, \textbf{84.6}, 0.1)\\
              ~& 2 1 1 1 2 & 8779 & 8793 & (0.0, 0.0, 0.0, 0.0, 0.0, 0.4, 0.4,\textbf{99.2}) & 11925 & 11975 & (0.0, 0.0, 0.0, 0.1, 0.0, \textbf{84.8}, \textbf{14.5}, 0.6) \\
              ~& 2 1 2 1 1 & 8843 & 8801 & (0.0, 0.2, \textbf{99.5}, 0.0, 0.0, 0.0, 0.0, 0.3)  & 11975 & 11992 & (\textbf{86.3}, \textbf{13.5}, 0.1, 0.1, 0.0, 0.0, 0.0, 0.0) \\
               & 2 1 2 1 2 & 8843 & 8842 & (0.0, 0.0, 0.0, 0.1 ,0.1, \textbf{48.7}, \textbf{51.1},0.0) & 11973 & 11993 &   (\textbf{13.5}, \textbf{86.5}, 0.0, 0.0, 0.0, 0.0, 0.0, 0.0)\\
              ~& 2 1 3 1 2 & 8793 & 8844 & (0.0, 0.0, 0.0, 0.0 ,0.1, \textbf{51.1}, \textbf{48.7},0.10) & 11936 & 12020 & (0.2, 0.0, \textbf{99.8}, 0.0, 0.0, 0.0, 0.0, 0.0)\\ \hline

\end{tabular}
\end{ruledtabular}
\end{table*}

Basing on these above analyses, we obtain the complete mass spectra of $\Omega_{ccb}$, $\Omega_{bbc}$, $\Omega_{ccc}$ and $\Omega_{bbb}$ baryons with quantum numbers up to $n = 4$ and $L = 4$. The results are listed in Tables \ref{massccb}-\ref{massbbb}. Many collaborations have focused on the mass spectra of these baryons with lower orbital excitations or radial excitations, which results are also listed in these two tables. In Ref. \cite{Yang:2019lsg}, Yang \emph{et al}. predicted the mass spectra of the triply baryons with quantum numbers up to $n = 2$ and $L = 2$, where the non-relativized quark model was adopted. In Ref. \cite{Silves:1996myf}, B. Silvestre-Brac employed the Faddeev formalism to predict the ground-state and lower excited state energies of triply baryons. From these tables, we can see that there is about $10\sim30$ MeV differences between our results and those in Refs. \cite{Yang:2019lsg,Silves:1996myf} for $\Omega_{ccb}$ and $\Omega_{bbc}$ system. As for the excited states of $\Omega_{ccc}$ and $\Omega_{bbb}$, the differences reach about $50\sim60$ MeV. Actually, if the dependence of results on model is considered, this mismatch is reasonable and acceptable. A similar study was performed in Ref. \cite{Serafin:2018aih}, where they applied the model of renormalization group procedure for effective particles (RGPEP). It is shown that the differences between our results and their predictions for $\Omega_{bbc}$, $\Omega_{ccc}$ and $\Omega_{bbb}$ are $10\sim30$ MeV. However, deviations reach more than 100 MeV for $\Omega_{ccb}$ baryons. S.-X. Qin \emph{et} \emph{al}. also reported their theoretical values which were obtained by Faddeev equation \cite{Qin:2019hgk}. It is obvious that their predicted masses are much lower than the results of other collaborations. In Ref. \cite{Flynn:2011gf}, the authors adopted the $\Delta$-shaped and $Y$-shaped potentials to investigate the ground state masses of triply heavy baryons. Their results are also presented in the last two columns in Tables \ref{massccb}-\ref{massbbb}. It is indicated that the masses obtained from $Y$-shaped potential are $30\sim50$ MeV higher than our results and those calculated by $\Delta$-shaped potential. Aa a verification, it will be interesting to study the excited state masses of the triply heavy baryons with $\Delta$ and $Y$-shaped potentials, which can also help to shed more light on the nature of the confinement potential in the baryon sector.

From Table \ref{massbbc}, another interesting characteristic about the orbital excited state of $\Omega_{bbc}$ baryon is shown. We can see that the mass of $1P(\frac{5}{2}^{-})$ state is 11562 MeV. This value is $40\sim60$ MeV higher than the other $P-$wave states. Besides, there also exist the similar feature for $1F(\frac{3}{2}^{-})$ state whose mass is 11921 MeV. It is $50\sim70$ MeV higher than the masses of other $F-$wave states. However, this phenomenon for $\Omega_{ccb}$ baryon is not so obvious as that of $\Omega_{bbc}$ system. Theoretically, baryons with the same orbital excitations should not have too much difference in there energies. To investigate this characteristic, all of the possible configurations about $1P(\frac{5}{2}^{-})$ and $1F(\frac{3}{2}^{-})$ are listed in Table \ref{ccb-bbc}. We can see that there only exist configuration with $\lambda$-mode ($l_{\rho}$,$l_{\lambda}$)=($0$,$1$) for $1P(\frac{5}{2}^{-})$ state in the allowed assignments of angular momentum. As for $1F(\frac{3}{2}^{-})$ state, only $\lambda$-mode and $\lambda$-$\rho$ mixing mode with ($l_{\rho}$,$l_{\lambda}$)=($0$,$3$), ($1$,$2$), and ($2$,$1$) are allowed, while $\rho$-mode ($l_{\rho}$,$l_{\lambda}$)=($3$,$0$) is forbidden. It has been indicated in Sec. \ref{IIIA} that the orbital excitations for $\Omega_{bbc}$ baryon are dominated by $\rho$-mode. Because of the disappearance of this orbitally excited mode, the lowest energies of $1P(\frac{5}{2}^{-})$ and $1F(\frac{3}{2}^{-})$ $\Omega_{bbc}$ baryons are much higher than those of other $P-$wave and $F-$wave states, respectively.

As for the uncertainties of the relativized quark model, it is very difficult for us to determine its exact value. It was claimed in Ref. \cite{GI1} that the uncertainties of constituent quark model depend on the quenched approximation and relativistic corrections. Considering these two effects, they claimed that the average accuracies are 25 MeV for light and heavy-light mesons and 10 MeV for heavy mesons, respectively. In our previous work \cite{GLY1}, the mass spectra of single heavy baryons were obtained by the relativized quark model. It was indicated that the deviations between predicted masses and measured ones are almost less than 20 MeV except for a few excited states. For doubly heavy baryons, the predicted mass for $\Xi_{cc}(\frac{1}{2}^{+})$ is 3640 MeV \cite{Yu:2022lel} which is about 19 MeV higher than experimental data $3621.4$ MeV. Basing on these previous analyses, we expect that the uncertainties of predicted masses of the triply heavy baryons are limited in 30 MeV.

\subsection{Regge trajectories of triply heavy baryons}

\begin{figure}[!h]
  \centering
   \subfigure[]{
   \begin{minipage}{4cm}
   \centering
   \includegraphics[width=4.5cm]{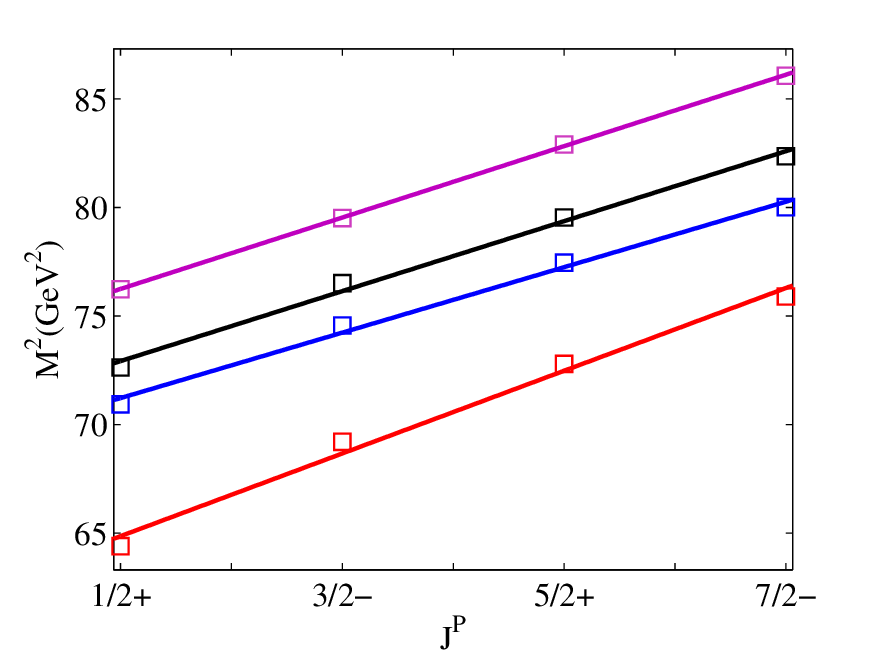}
  \end{minipage}
  }
 \subfigure[]{
   \begin{minipage}{4cm}
   \centering
   \includegraphics[width=4.5cm]{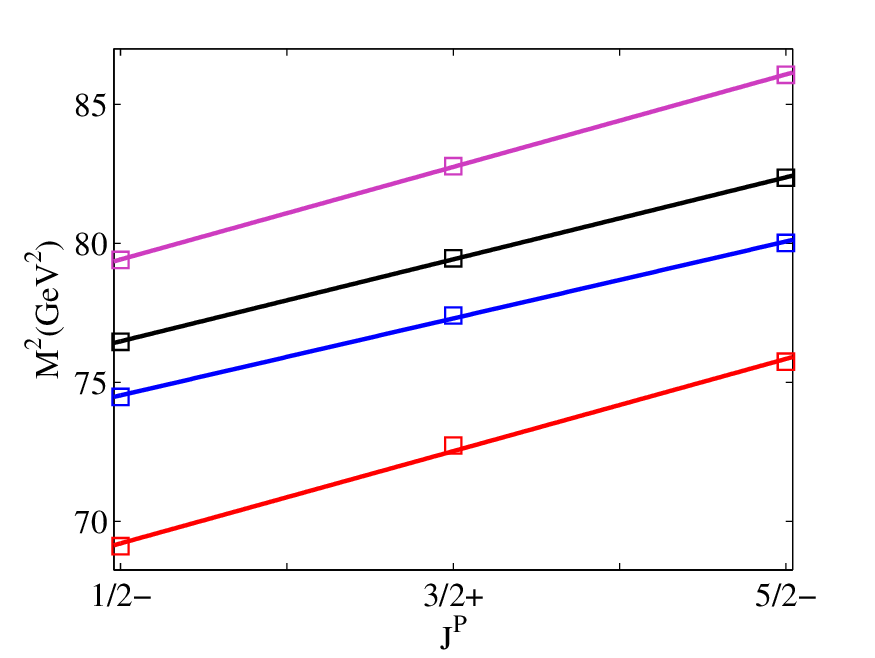}
  \end{minipage}
  }
  \caption{Parent and daughter($J$,$M^2$) Regge trajectories for $\Omega_{ccb}$ baryons with natural (a) and
 unnatural (b) parities.}
\label{traj1}
\end{figure}
\begin{figure}[!h]
  \centering
   \subfigure[]{
   \begin{minipage}{4cm}
   \centering
   \includegraphics[width=4.5cm]{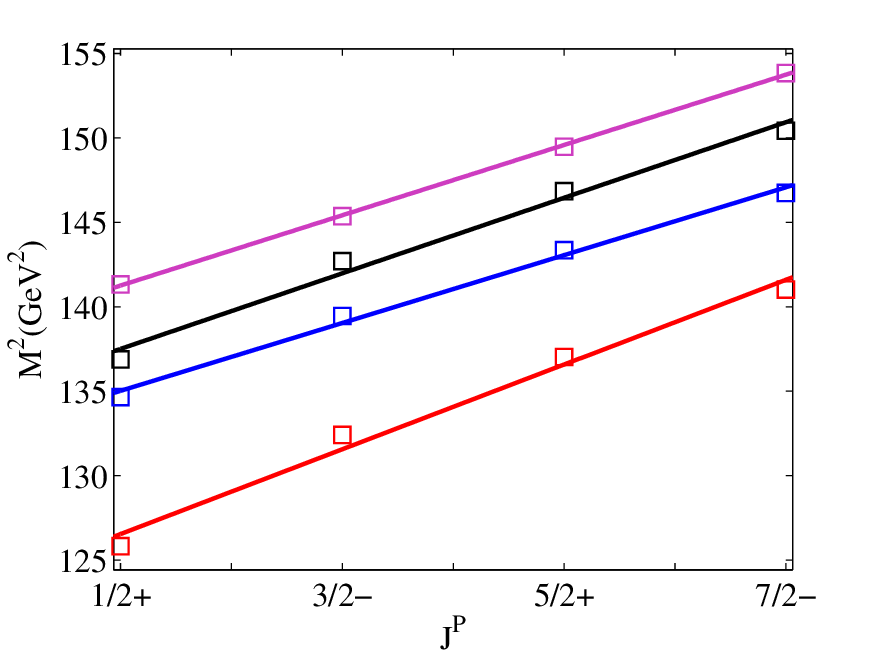}
  \end{minipage}
  }
 \subfigure[]{
   \begin{minipage}{4cm}
   \centering
   \includegraphics[width=4.5cm]{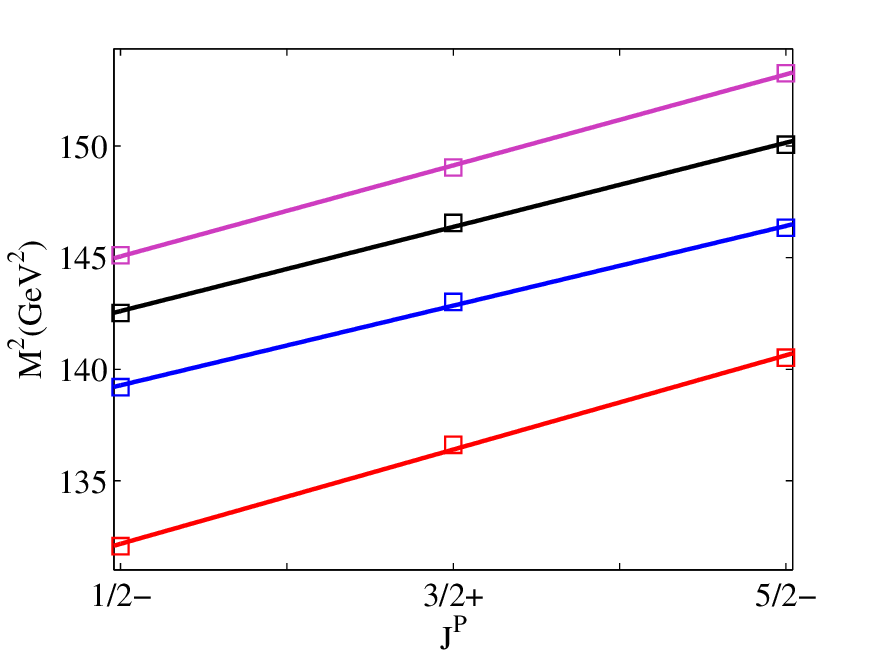}
  \end{minipage}
  }
  \caption{Same as in Fig. 6 but for $\Omega_{bbc}$ baryons.}
\label{traj2}
\end{figure}
\begin{figure}[!h]
  \centering
   \subfigure[]{
   \begin{minipage}{4cm}
   \centering
   \includegraphics[width=4.5cm]{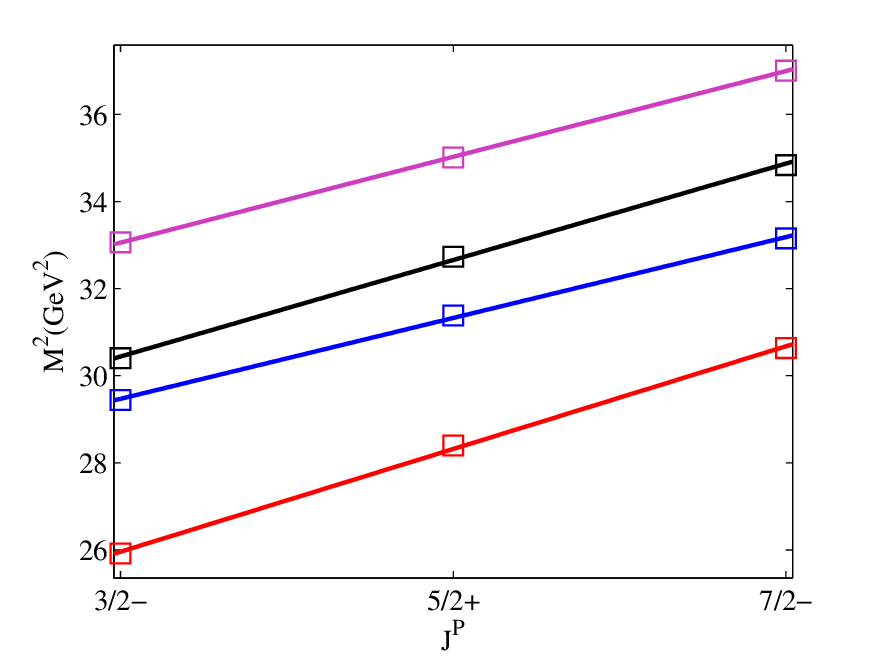}
  \end{minipage}
  }
 \subfigure[]{
   \begin{minipage}{4cm}
   \centering
   \includegraphics[width=4.5cm]{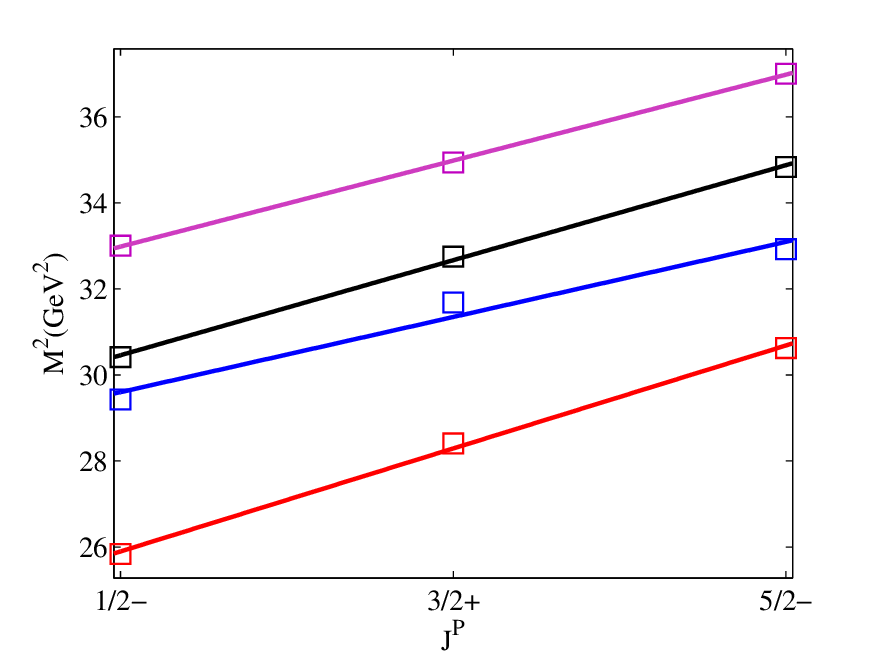}
  \end{minipage}
  }
  \caption{Same as in Fig. 6 but for $\Omega_{ccc}$ baryons.}
\label{traj3}
\end{figure}
\begin{figure}[!h]
  \centering
   \subfigure[]{
   \begin{minipage}{4cm}
   \centering
   \includegraphics[width=4.5cm]{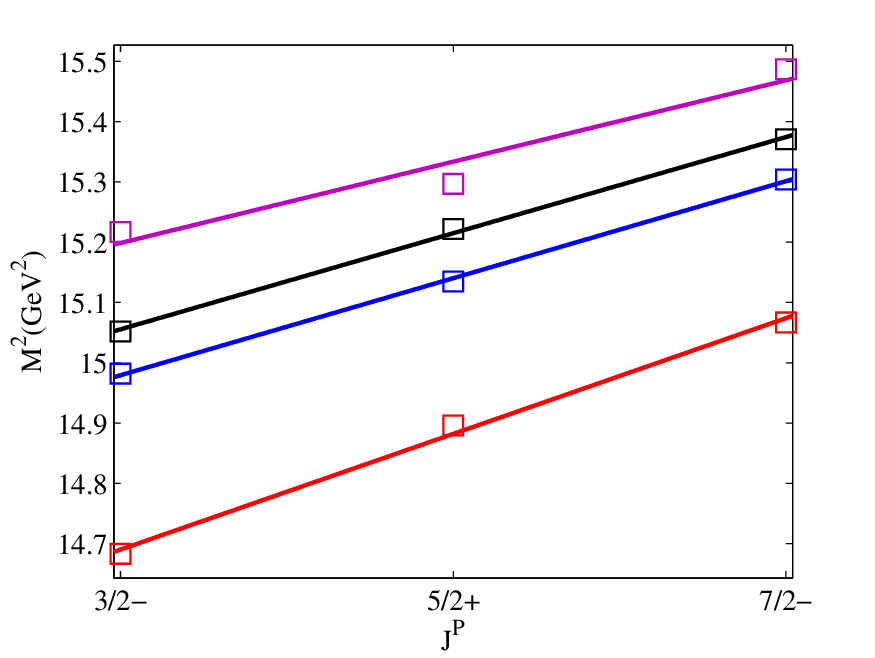}
  \end{minipage}
  }
 \subfigure[]{
   \begin{minipage}{4cm}
   \centering
   \includegraphics[width=4.5cm]{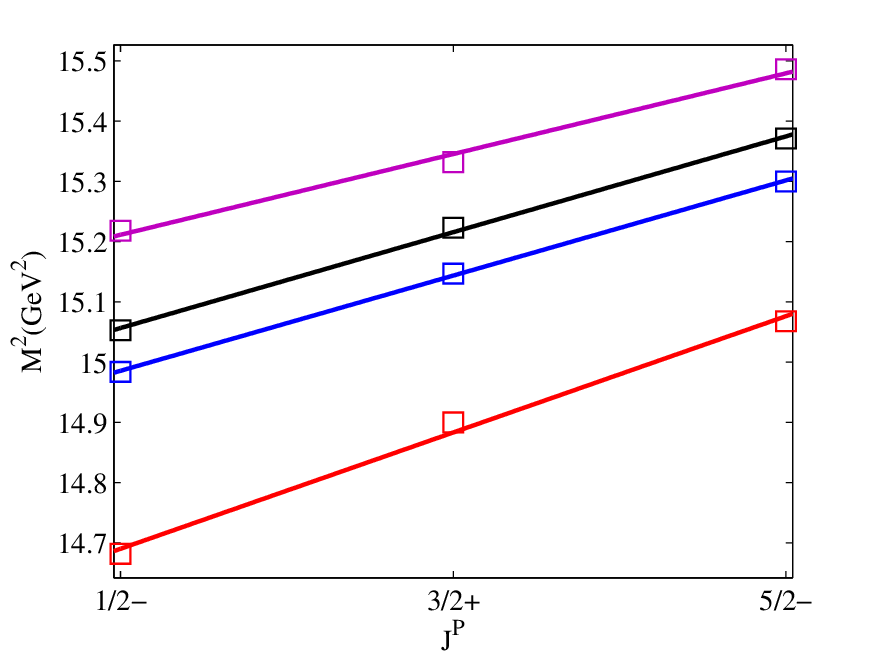}
  \end{minipage}
  }
  \caption{Same as in Fig. 6 but for $\Omega_{bbb}$ baryons.}
\label{traj4}
\end{figure}

The Regge theory which was first proposed by T. Regge in 1959 \cite{Regge1,Regge2} is very successful in describing mass spectra of the hadrons \cite{Regge3,Regge4,Regge5,Regge6,Regge7,Regge8,Regge9,Regge10,Ebert,Guo:2008he}.
In our previous work, we successfully constructed the Regge trajectories for the single and doubly heavy baryons \cite{GLY1,Yu:2022lel,ZYL1}.
In the present work, we successfully obtain the complete mass spectra of the $1S\sim4S$, $1P\sim4P$, $1D\sim4D$, and $1F\sim4F$ state for triply heavy baryons. This makes it easy for us to construct their Regge trajectories in ($J$,$M^{2}$) plane.
The triply heavy baryons can be classified into two groups which have natural parity $P=(-1)^{J-\frac{1}{2}}$ and unnatural parity $P=(-1)^{J+\frac{1}{2}}$. The Regge trajectory in the ($J$,$M^{2}$) plane is defined as,
\begin{eqnarray}
M^{2}=\alpha J+\alpha_{0}
\end{eqnarray}
where $\alpha$ and $\alpha_{0}$ are slope and intercept. Using this above equation, we obtain the Regge trajectories of $\Omega_{ccb}$, $\Omega_{bbc}$, $\Omega_{ccc}$ and $\Omega_{bbb}$ baryons which are shown in Figs. \ref{traj1}$-$\ref{traj4} respectively. In these figures, the predicted masses with quark model are denoted by squares. The ground and radial excited states are plotted from bottom to top.

\begin{table*}[htbp]
\begin{ruledtabular}\caption{Fitted parameters $\alpha$ and $\alpha_{0}$ for the slope and intercept of the ($J$,$M^2$) parent and daughter Regge trajectories for triply heavy baryons.}
\label{fitpara}
\begin{tabular}{c c c| c c}
Trajectory&$\alpha$(Gev$^{2}$)&$\alpha_{0}$(Gev$^{2}$) &$\alpha$(Gev$^{2}$)&$\alpha_{0}$(Gev$^{2}$) \\ \hline
&$\Omega_{ccb}(\frac{1}{2}^{+})$&~&$\Omega_{ccb}(\frac{1}{2}^{-})$&~\\
parent&3.81$\pm$0.81&62.96$\pm$2.70&3.32$\pm$1.10&67.55$\pm$4.12\\
1 daughter&3.01$\pm$0.75&69.71$\pm$1.50&2.77$\pm$0.90&73.15$\pm$2.15\\
2 daughter&3.22$\pm$0.75&71.31$\pm$1.70&2.95$\pm$0.45&75.00$\pm$0.63\\
3 daughter&3.29$\pm$0.15&74.6$\pm$0.35&3.33$\pm$0.36&77.75$\pm$0.52\\ \hline
&$\Omega_{bbc}(\frac{1}{2}^{+})$&~&$\Omega_{bbc}(\frac{1}{2}^{-})$&~\\
parent&5.02$\pm$1.82&124.00$\pm$4.52&4.23$\pm$2.32&130.10$\pm$4.17\\
1 daughter&4.02$\pm$0.95&133.00$\pm$2.55&3.57$\pm$1.83&137.50$\pm$3.20\\
2 daughter&4.48$\pm$1.58&135.30$\pm$3.59&3.77$\pm$1.95&140.70$\pm$3.54\\
3 daughter&4.16$\pm$0.22&139.20$\pm$0.50&4.08$\pm$1.14&143.00$\pm$1.90\\ \hline
&$\Omega_{ccc}(\frac{3}{2}^{-})$&~&$\Omega_{ccc}(\frac{1}{2}^{-})$&~\\
parent&2.36$\pm$0.95&22.42$\pm$2.32&2.39$\pm$0.83&24.70$\pm$2.13\\
1 daughter&1.86$\pm$0.65&26.68$\pm$1.63&1.75$\pm$3.62&28.73$\pm$6.81\\
2 daughter&2.22$\pm$0.82&27.12$\pm$2.07&2.21$\pm$0.98&29.35$\pm$1.62\\
3 daughter&1.97$\pm$0.16&30.10$\pm$0.48&2.00$\pm$0.50&31.98$\pm$0.62\\ \hline
&$\Omega_{bbb}(\frac{3}{2}^{-})$&~&$\Omega_{bbb}(\frac{1}{2}^{-})$&~\\
parent&0.19$\pm$0.15&14.40$\pm$0.40&0.19$\pm$0.18&14.59$\pm$0.37\\
1 daughter&0.16$\pm$0.06&14.74$\pm$0.15&0.16$\pm$0.04&14.91$\pm$0.05\\
2 daughter&0.16$\pm$0.08&14.82$\pm$0.42&0.16$\pm$0.08&14.98$\pm$0.16\\
3 daughter&0.14$\pm$0.21&15.00$\pm$1.52&0.13$\pm$0.87&15.14$\pm$0.27\\
\end{tabular}
\end{ruledtabular}
\end{table*}
The straight lines in these figures are obtained by linear fitting of the numerical results. The fitted slopes and intercepts of the Regge trajectories are listed in Table \ref{fitpara}. It can be seen that all of the predicted masses in the present work are fitted nicely into linear trajectories on the ($J$,$M^{2}$) plane. These results can help us to assign an accurate position in the mass spectra for experimentally observed $\Omega_{ccb}$ and $\Omega_{bbc}$ baryons in the future.

\section{ Conclusions}

In this work, we have systematically investigate the mass spectra, the r.m.s. radii and the radial density distributions of the $\Omega_{ccb}$ with $\lambda$-mode and $\Omega_{bbc}$ with $\rho$-mode in the frame work of relativized quark model. All parameters used in present work such as quark masses and inter-quark potentials in the Hamiltonian are consistent with those of our previous work\cite{GLY1}. According to analyzing the excited energies of different orbitally excited modes, we find that the dominant orbital excitations are associated with the heavier quark in charmed-bottom baryons. This characteristic is consistent well with our previous conclusion which is named as the mechanism of heavy quark dominance \cite{Li:2023gbo}. In addition, we also find that the lowest energy level is further lowered by configuration mixing of different angular momentum assignments. Basing on these analyses, the complete mass spectra of the ground, orbitally and radially excited states($1S\sim4S$, $1P\sim4P$, $1D\sim4D$, $1F\sim4F$ and $1G\sim4G$) of triply heavy baryons are systematically studied(Tables \ref{massccb}-\ref{massbbb}). Finally, with the predicted mass spectra, we also construct the Regge trajectories in ($J$,$M^{2}$) plane.

Up to now, no experimental data related to $\Omega_{ccb}$, $\Omega_{bbc}$, $\Omega_{ccc}$ and $\Omega_{bbb}$ triply heavy baryons are reported. For most theoretical researches, only masses of the ground state, lower radially and orbitally excited states are explored. If model uncertainties are considered, our predicted results are comparable with some of the results \cite{Yang:2019lsg,Silves:1996myf}. In summary, we hope these analyses will be helpful to search for triply heavy baryons in future experiments.

\begin{Large}
Acknowledgments
\end{Large}
This project is supported by National Natural Science Foundation, Grant Number 12175068 and Natural Science Foundation of HeBei Province, Grant Number A2024502002.

\begin{table*}[htbp]
\begin{ruledtabular}\caption{Predicted masses(in MeV) and r.m.s. radii(in fm) of the $\lambda-$mode $\Omega_{ccb}$ baryons for different configurations}
\label{ccbLambda}
\begin{tabular}{c c c c c |c c c c c}
			$l_{\rho}$  $l_{\lambda}$ L s j & $nL$($J^{P}$)  &$\sqrt {\langle {r_{\rho}^{2}}\rangle }$ &$\sqrt {\langle {r_{\lambda}^{2}}\rangle }$&M &$l_{\rho}$  $l_{\lambda}$ L s j & $nL$($J^{P}$)  &$\sqrt {\langle {r_{\rho}^{2}}\rangle }$ &$\sqrt {\langle {r_{\lambda}^{2}}\rangle }$&M \\ \hline
			\multirow{4}{*}{0 0 0 1 1 }
			~& $1S$($\frac{1}{2}^{+}$) & 0.387 & 0.285 & 8025  & \multirow{4}{*}{0 2 2 1 2}
			~ & $1D$($\frac{3}{2}^{+}$) & 0.455 & 0.572 & 8528   \\
			~& $2S$($\frac{1}{2}^{+}$) & 0.527 & 0.509 & 8422   &~& $2D$($\frac{3}{2}^{+}$) & 0.486 & 0.872 & 8798  \\
			~& $3S$($\frac{1}{2}^{+}$) & 0.664 & 0.450 & 8522  &~ & $3D$($\frac{3}{2}^{+}$) & 0.813 & 0.624 & 8914   \\
			~& $4S$($\frac{1}{2}^{+}$) & 0.583 & 0.710 & 8731  &~ & $4D$($\frac{3}{2}^{+}$) & 0.594 & 0.938 & 9104  \\ \hline	
			\multirow{4}{*}{0 0 0 1 1}
			~ & $1S$($\frac{3}{2}^{+}$) & 0.393 & 0.297 & 8046  &\multirow{4}{*}{0 2 2 1 2}
			~ & $1D$($\frac{5}{2}^{+}$) & 0.456 & 0.577 & 8532  \\
			~ & $2S$($\frac{3}{2}^{+}$) & 0.527 & 0.525 & 8438 &~ & $2D$($\frac{5}{2}^{+}$) & 0.486 & 0.876 & 8801   \\
			~ & $3S$($\frac{3}{2}^{+}$) & 0.673 & 0.454 & 8563  &~ & $3D$($\frac{5}{2}^{+}$) & 0.815 & 0.627 & 8918  \\
			~ & $4S$($\frac{3}{2}^{+}$) & 0.579 & 0.719 & 8745  &~ & $4D$($\frac{5}{2}^{+}$) & 0.595 & 0.939 & 9105   \\ \hline
			\multirow{4}{*}{0 1 1 1 0}
			~ & $1P$($\frac{1}{2}^{-}$) & 0.434 & 0.440 & 8317  &	\multirow{4}{*}{0 2 2 1 3}
			~ & $1D$($\frac{5}{2}^{+}$) & 0.455 & 0.572 & 8527  \\
			~ & $2P$($\frac{1}{2}^{-}$) & 0.498 & 0.710 & 8633  &~ & $2D$($\frac{5}{2}^{+}$) & 0.486 & 0.872 & 8798 \\
			~ & $3P$($\frac{1}{2}^{-}$) & 0.757 & 0.533 & 8746 &~ & $3D$($\frac{5}{2}^{+}$) & 0.813 & 0.624 & 8914  \\
			~ & $4P$($\frac{1}{2}^{-}$) & 0.546 & 0.808 & 8915  &~ & $4D$($\frac{5}{2}^{+}$) & 0.583 & 0.923 & 9098    \\ \hline
			\multirow{4}{*}{0 1 1 1 1}
			~ & $1P$($\frac{1}{2}^{-}$) & 0.433 & 0.437 & 8313  & \multirow{4}{*}{0 2 2 1 3}
			~ & $1D$($\frac{7}{2}^{+}$) & 0.456 & 0.577 & 8532  \\
			~ & $2P$($\frac{1}{2}^{-}$) & 0.498 & 0.706 & 8630   &~ & $2D$($\frac{7}{2}^{+}$) & 0.486 & 0.876 & 8802  \\
			~ & $3P$($\frac{1}{2}^{-}$) & 0.755 & 0.532 & 8744  &~ & $3D$($\frac{7}{2}^{+}$) & 0.815 & 0.627 & 8918   \\
			~ & $4P$($\frac{1}{2}^{-}$) & 0.545 & 0.805 & 8911  &~ & $4D$($\frac{7}{2}^{+}$) & 0.596 & 0.941 & 9106
			\\ \hline
			\multirow{4}{*}{0 1 1 1 1}
			~ & $1P$($\frac{3}{2}^{-}$) & 0.435 & 0.441 & 8319  &\multirow{4}{*}{0 3 3 1 2}
			~ & $1F$($\frac{3}{2}^{-}$) & 0.466 & 0.699 & 8707       \\
			~ & $2P$($\frac{3}{2}^{-}$) & 0.497 & 0.712 & 8635   &~ & $2F$($\frac{3}{2}^{-}$) & 0.483 & 0.992 & 8942 \\
			~ & $3P$($\frac{3}{2}^{-}$) & 0.758 & 0.533 & 8747   &~ & $3F$($\frac{3}{2}^{-}$) & 0.849 & 0.734 & 9071  \\
			~ & $4P$($\frac{3}{2}^{-}$) & 0.546 & 0.810 & 8917  &	~ & $4F$($\frac{3}{2}^{-}$) & 0.837 & 1.049 & 9273 	\\ \hline
			\multirow{4}{*}{0 1 1 1 2}
			~ & $1P$($\frac{3}{2}^{-}$) & 0.433 & 0.435 & 8311   &\multirow{4}{*}{0 3 3 1 2}
			~ & $1F$($\frac{5}{2}^{-}$) & 0.466 & 0.702 & 8709   \\
			~ & $2P$($\frac{3}{2}^{-}$) & 0.498 & 0.704 & 8629   &~ & $2F$($\frac{5}{2}^{-}$) & 0.483 & 0.992 & 8943  \\
			~ & $3P$($\frac{3}{2}^{-}$) & 0.754 & 0.531 & 8742  &~ & $3F$($\frac{5}{2}^{-}$) & 0.850 & 0.737 & 9073   \\
			~ & $4P$($\frac{3}{2}^{-}$) & 0.544 & 0.804 & 8908   &~ & $4F$($\frac{5}{2}^{-}$) & 0.847 & 1.050 & 9275
			\\ \hline
			\multirow{4}{*}{0 1 1 1 2}
			~ & $1P$($\frac{5}{2}^{-}$) & 0.435 & 0.443 & 8321   &\multirow{4}{*}{0 3 3 1 3}
			~ & $1F$($\frac{5}{2}^{-}$) & 0.466 & 0.699 & 8707   \\
			~ & $2P$($\frac{5}{2}^{-}$) & 0.497 & 0.714 & 8637   &~ & $2F$($\frac{5}{2}^{-}$) & 0.483 & 0.992 & 8942 \\
			~ & $3P$($\frac{5}{2}^{-}$) & 0.759 & 0.534 & 8749   &~ & $3F$($\frac{5}{2}^{-}$) & 0.849 & 0.734 & 9071   \\
			~ & $4P$($\frac{5}{2}^{-}$) & 0.547 & 0.812 & 8919 &~ & $4F$($\frac{5}{2}^{-}$) & 0.837 & 1.049 & 9273
			\\ \hline
			\multirow{4}{*}{0 2 2 1 1}
			~ & $1D$($\frac{1}{2}^{+}$) & 0.455 & 0.572 & 8527  &\multirow{4}{*}{0 3 3 1 3}
			~ & $1F$($\frac{7}{2}^{-}$) & 0.466 & 0.702 & 8710   \\
			~ & $2D$($\frac{1}{2}^{+}$) & 0.486 & 0.872 & 8798  & &$2F$($\frac{7}{2}^{-}$) & 0.483 & 0.992 & 8943 \\
			~ & $3D$($\frac{1}{2}^{+}$) & 0.813 & 0.624 & 8914  & &$3F$($\frac{7}{2}^{-}$) & 0.850 & 0.737 & 9073   \\
			~ & $4D$($\frac{1}{2}^{+}$) & 0.583 & 0.923 & 9098  & &$4F$($\frac{7}{2}^{-}$) & 0.847 & 1.050 & 9275  \\   \hline
			 \multirow{4}{*}{0 2 2 1 1}
			~ & $1D$($\frac{3}{2}^{+}$) & 0.456 & 0.576 & 8531  &\multirow{4}{*}{0 3 3 1 4}
			~ & $1F$($\frac{7}{2}^{-}$) & 0.466 & 0.699 & 8707   \\
			~& $2D$($\frac{3}{2}^{+}$) & 0.486 & 0.875 & 8801  &&$2F$($\frac{7}{2}^{-}$) & 0.483 & 0.992 & 8941 \\
			~ & $3D$($\frac{3}{2}^{+}$) & 0.815 & 0.627 & 8917 & &$3F$($\frac{7}{2}^{-}$) & 0.849 & 0.733 & 9071 \\
			~ & $4D$($\frac{3}{2}^{+}$) & 0.594 & 0.938 & 9104 & &$4F$($\frac{7}{2}^{-}$) & 0.837 & 1.049 & 9272 \\
\end{tabular}
\end{ruledtabular}
\end{table*}
\begin{table}[htbp]
\begin{ruledtabular}\caption{Predicted masses(in MeV) and r.m.s. radii(in fm) of the $\rho-$mode $\Omega_{bbc}$ baryons for different configurations}
\label{bbcRho}
\begin{tabular}{c c c c c}
			$l_{\rho}$  $l_{\lambda}$ L s j & $nL$($J^{P}$)  &$\sqrt {\langle {r_{\rho}^{2}}\rangle }$ &$\sqrt {\langle {r_{\lambda}^{2}}\rangle }$&M   \\ \hline
			\multirow{4}{*}{0 0 0 1 1 }
			~& $1S$($\frac{1}{2}^{+}$) & 0.272 & 0.297 & 11217  \\
			~& $2S$($\frac{1}{2}^{+}$) & 0.506 & 0.391 & 11604  \\
			~& $3S$($\frac{1}{2}^{+}$) & 0.346 & 0.584 & 11700  \\
			~& $4S$($\frac{1}{2}^{+}$) & 0.722 & 0.432 & 11888  \\ \hline
			\multirow{4}{*}{0 0 0 1 1}
			~ & $1S$($\frac{3}{2}^{+}$) & 0.275 & 0.307 & 11236  \\
			~ & $2S$($\frac{3}{2}^{+}$) & 0.512 & 0.398 & 11617  \\
			~ & $3S$($\frac{3}{2}^{+}$) & 0.345 & 0.593 & 11709   \\
			~ & $4S$($\frac{3}{2}^{+}$) & 0.724 & 0.436 & 11899  \\ \hline
			\multirow{4}{*}{1 0 1 0 1}
			~ & $1P$($\frac{1}{2}^{-}$) & 0.420 & 0.329 & 11492  \\
			~ & $2P$($\frac{1}{2}^{-}$) & 0.672 & 0.396 & 11798  \\
			~ & $3P$($\frac{1}{2}^{-}$) & 0.485 & 0.627 & 11938  \\
			~ & $4P$($\frac{1}{2}^{-}$) & 0.771 & 0.438 & 12046   \\ \hline
			\multirow{4}{*}{1 0 1 0 1}
			~ & $1P$($\frac{3}{2}^{-}$) & 0.423 & 0.337 & 11507  \\
			~ & $2P$($\frac{3}{2}^{-}$) & 0.679 & 0.403 & 11809  \\
			~ & $3P$($\frac{3}{2}^{-}$) & 0.485 & 0.635 & 11946  \\
			~ & $4P$($\frac{3}{2}^{-}$) & 0.771 & 0.443 & 12057  \\ \hline
			\multirow{4}{*}{2 0 2 1 1}
			~ & $1D$($\frac{1}{2}^{+}$) & 0.543 & 0.355 & 11690  \\
			~ & $2D$($\frac{1}{2}^{+}$) & 0.832 & 0.421 & 11960   \\
			~ & $3D$($\frac{1}{2}^{+}$) & 0.610 & 0.650 & 12107  \\
			~ & $4D$($\frac{1}{2}^{+}$) & 0.786 & 0.455 & 12209  \\ \hline
			\multirow{4}{*}{2 0 2 1 1}
			~ & $1D$($\frac{3}{2}^{+}$) & 0.547 & 0.363 & 11700  \\
			~ & $2D$($\frac{3}{2}^{+}$) & 0.838 & 0.429 & 11969     \\
			~ & $3D$($\frac{3}{2}^{+}$) & 0.610 & 0.658 & 12114  \\
			~ & $4D$($\frac{3}{2}^{+}$) & 0.788 & 0.461 & 12219  \\ \hline
			\multirow{4}{*}{2 0 2 1 2}
			~ & $1D$($\frac{3}{2}^{+}$) & 0.545 & 0.353 & 11688  \\
			~ & $2D$($\frac{3}{2}^{+}$) & 0.834 & 0.419 & 11959  \\
			~ & $3D$($\frac{3}{2}^{+}$) & 0.612 & 0.648 & 12106  \\
			~ & $4D$($\frac{3}{2}^{+}$) & 0.786 & 0.454 & 12208 \\ \hline
			\multirow{4}{*}{2 0 2 1 2}
			~ & $1D$($\frac{5}{2}^{+}$) & 0.550 & 0.366 & 11706  \\
			~ & $2D$($\frac{5}{2}^{+}$) & 0.843 & 0.431 & 11973  \\
			~ & $3D$($\frac{5}{2}^{+}$) & 0.612 & 0.661 & 12118  \\
			~ & $4D$($\frac{5}{2}^{+}$) & 0.789 & 0.465 & 12226  \\ \hline
			\multirow{4}{*}{2 0 2 1 3}
			~ & $1D$($\frac{5}{2}^{+}$) & 0.547 & 0.351 & 11688  \\
			~ & $2D$($\frac{5}{2}^{+}$) & 0.838 & 0.418 & 11959  \\
			~ & $3D$($\frac{5}{2}^{+}$) & 0.616 & 0.646 & 12107   \\
			~ & $4D$($\frac{5}{2}^{+}$) & 0.787 & 0.455 & 12211   \\ \hline
			\multirow{4}{*}{2 0 2 1 3}
			~ & $1D$($\frac{7}{2}^{+}$) & 0.555 & 0.369 & 11713  \\
			~ & $2D$($\frac{7}{2}^{+}$) & 0.849 & 0.434 & 11979   \\
			~ & $3D$($\frac{7}{2}^{+}$) & 0.615 & 0.664 & 12123  \\
			~ & $4D$($\frac{7}{2}^{+}$) & 0.794 & 0.471 & 12237  \\ \hline
			\multirow{4}{*}{3 0 3 0 3}
			~ & $1F$($\frac{5}{2}^{-}$) & 0.655 & 0.373 & 11854   \\
			~ & $2F$($\frac{5}{2}^{-}$) & 0.980 & 0.441 & 12097   \\
			~ & $3F$($\frac{5}{2}^{-}$) & 0.727 & 0.665 & 12250  \\
			~ & $4F$($\frac{5}{2}^{-}$) & 0.865 & 0.531 & 12380  \\ \hline
			\multirow{4}{*}{3 0 3 0 3}
			~ & $1F$($\frac{7}{2}^{-}$) & 0.662 & 0.390 & 11875   \\
			~ & $2F$($\frac{7}{2}^{-}$) & 0.984 & 0.456 & 12114   \\
			~ & $3F$($\frac{7}{2}^{-}$) & 0.729 & 0.681 & 12265  \\
			~ & $4F$($\frac{7}{2}^{-}$) & 0.891 & 0.562 & 12403
\end{tabular}
\end{ruledtabular}
\end{table}


\begin{thebibliography}{1}
\bibitem{ParticleDataGroup:2024cfk}
S.~Navas \textit{et al.} [Particle Data Group],
``Review of particle physics,''
\href{https://doi:10.1103/PhysRevD.110.030001}{Phys. Rev. D \textbf{110}, 030001 (2024)}.

\bibitem{LHCb:2017iph}
R.~Aaij \textit{et al.} [LHCb],
``Observation of the doubly charmed baryon $\Xi_{cc}^{++}$,''
\href{https://doi:10.1103/PhysRevLett.119.112001}{Phys. Rev. Lett. \textbf{119}, 112001 (2017)}.

\bibitem{GomshiNobary:2003sf}
M.~A.~Gomshi Nobary,
``Fragmentation production of $\Omega_{ccc}$ and $\Omega_{bbb}$ baryons,''
\href{https://doi:10.1016/j.physletb.2002.12.001}{Phys. Lett. B \textbf{559}, 239-244 (2003)}.

\bibitem{GomshiNobary:2004mq}
M.~A.~Gomshi Nobary and R.~Sepahvand,
``Fragmentation of triply heavy baryons,''
\href{https://doi:10.1103/PhysRevD.71.034024}{Phys. Rev. D \textbf{71}, 034024 (2005)}.

\bibitem{GomshiNobary:2005ur}
M.~A.~Gomshi Nobary and R.~Sepahvand,
``An Ivestigation of triply heavy baryon production at hadron colliders,''
\href{https://doi:10.1016/j.nuclphysb.2006.01.043}{Nucl. Phys. B \textbf{741}, 34 (2006)}.

\bibitem{He:2014tga}
H.~He, Y.~Liu and P.~Zhuang,
``$\Omega_{ccc}$ production in high energy nuclear collisions,''
\href{https://doi:10.1016/j.physletb.2015.04.049}{Phys. Lett. B \textbf{746}, 59 (2015)}.

\bibitem{Zhao:2017gpq}
J.~Zhao and P.~Zhuang,
``Multicharmed Baryon Production in High Energy Nuclear Collisions,''
\href{https://doi:10.1007/s00601-017-1255-9}{Few Body Syst. \textbf{58}, 100 (2017)}.

\bibitem{Baranov:2004er}
S.~P.~Baranov and V.~L.~Slad,
``Production of triply charmed Omega(ccc) baryons in $e^+ e^-$ annihilation,''
\href{https://doi:10.1134/1.1707141}{Phys. Atom. Nucl. \textbf{67}, 808 (2004)}.

\bibitem{Chen:2011mb}
Y.~Q.~Chen and S.~Z.~Wu,
``Production of Triply Heavy Baryons at LHC,''
\href{https://doi:10.1007/JHEP08(2011)144}{JHEP \textbf{08}, 144 (2011)}.

\bibitem{Huang:2021jxt}
F.~Huang, J.~Xu and X.~R.~Zhang,
``Deciphering weak decays of triply heavy baryons by SU(3) analysis,''
\href{https://doi:10.1140/epjc/s10052-021-09729-x}{Eur. Phys. J. C \textbf{81}, 976 (2021)}.

\bibitem{Wang:2022ias}
W.~Wang and Z.~P.~Xing,
``Weak decays of triply heavy baryons in light front approach,''
\href{https://doi:10.1016/j.physletb.2022.137402}{Phys. Lett. B \textbf{834}, 137402 (2022)}.

\bibitem{Zhao:2022vfr}
Z.~X.~Zhao and Q.~Yang,
``Weak decays of triply heavy baryons in the light-front approach,''
arXiv:2204.00759 [hep-ph]

\bibitem{Lu:2024bqw}
F.~Lu, H.~W.~Ke and X.~H.~Liu,
``Weak decays of the triply heavy baryons in the three-quark picture with the light-front quark model,''
\href{https://doi:10.1140/epjc/s10052-024-12732-7}{Eur. Phys. J. C \textbf{84}, 452 (2024)}.

\bibitem{Hasenfratz:1980ka}
P.~Hasenfratz, R.~R.~Horgan, J.~Kuti and J.~M.~Richard,
``Heavy Baryon Spectroscopy in the {QCD} Bag Model,''
\href{https://doi:10.1016/0370-2693(80)90906-5}{Phys. Lett. B \textbf{94}, 401 (1980)}.

\bibitem{Bag2}
A.~Bernotas and V.~Simonis,
``Heavy hadron spectroscopy and the bag model,''
\href{https://doi.org/10.3952/lithjphys.49110}{Lith. J. Phys. \textbf{49}, 19 (2009)}.

\bibitem{Patel:2008mv}
B.~Patel, A.~Majethiya and P.~C.~Vinodkumar,
``Masses and Magnetic moments of Triply Heavy Flavour Baryons in Hypercentral Model,''
\href{https://doi:10.1007/s12043-009-0061-4}{Pramana \textbf{72}, 679 (2009)}.

\bibitem{Shah:2017jkr}
Z.~Shah and A.~K.~Rai,
``Masses and Regge trajectories of triply heavy $\Omega_{ccc}$ and $\Omega_{bbb}$ baryons,''
\href{https://doi:10.1140/epja/i2017-12386-2}{Eur. Phys. J. A \textbf{53}, 195 (2017)}.

\bibitem{Shah:2018div}
Z.~Shah and A.~K.~Rai,
``Ground and Excited State Masses of the $\Omega _{bbc}$ Baryon,''
\href{https://doi:10.1007/s00601-018-1398-3}{Few Body Syst. \textbf{59}, 76 (2018)}.

\bibitem{Shah:2018bnr}
Z.~Shah and A.~Kumar Rai,
``Spectroscopy of the $\Omega_{ccb}$ baryon in the hypercentral constituent quark model,''
\href{https://doi:10.1088/1674-1137/42/5/053101}{Chin. Phys. C \textbf{42}, 053101 (2018)}.

\bibitem{Liu:2019vtx}
M.~S.~Liu, Q.~F.~L\"u and X.~H.~Zhong,
``Triply charmed and bottom baryons in a constituent quark model,''
\href{https://doi:10.1103/PhysRevD.101.074031}{Phys. Rev. D \textbf{101}, 074031 (2020)}.

\bibitem{Yang:2019lsg}
G.~Yang, J.~Ping, P.~G.~Ortega and J.~Segovia,
``Triply heavy baryons in the constituent quark model,''
\href{https://doi:10.1088/1674-1137/44/2/023102}{Chin. Phys. C \textbf{44}, 023102 (2020)}.

\bibitem{Migura:2006ep}
S.~Migura, D.~Merten, B.~Metsch and H.~R.~Petry,
``Charmed baryons in a relativistic quark model,''
\href{https://doi:10.1140/epja/i2006-10017-9}{Eur. Phys. J. A \textbf{28}, 41 (2006)}.

\bibitem{Martynenko:2007je}
A.~P.~Martynenko,
``Ground-state triply and doubly heavy baryons in a relativistic three-quark model,''
\href{https://doi:10.1016/j.physletb.2008.04.030}{Phys. Lett. B \textbf{663}, 317 (2008)}.

\bibitem{Silves:1996myf}
B.~Silvestre-Brac,
``Spectrum and static properties of heavy baryons,''
\href{https://doi:10.1007/s006010050028}{Few Body Syst. \textbf{20}, 1 (1996)}.

\bibitem{Jia:2006gw}
Y.~Jia,
``Variational study of weakly coupled triply heavy baryons,''
\href{https://doi:10.1088/1126-6708/2006/10/073}{JHEP \textbf{10}, 073 (2006)}.

\bibitem{QM0}
W.~Roberts and M.~Pervin,
``Heavy baryons in a quark model,''
\href{https://doi.org/10.1142/S0217751X08041219}{Int. J. Mod. Phys. A \textbf{23}, 2817 (2008)}.

\bibitem{QM18}
Z.~Ghalenovi, A.~A.~Rajabi and M.~Hamzavi,
``The heavy baryon masses in variational approach and spin-isospin dependence,''
\href{https://doi.org/10.5506/APhysPolB.42.1849}{Acta Phys. Polon. B \textbf{42}, 1849 (2011)}.

\bibitem{Shah:2019jxp}
Z.~Shah and A.~K.~Rai,
``Mass spectra of triply heavy charm-beauty baryons,''
\href{https://doi:10.1051/epjconf/201920206001}{EPJ Web Conf. \textbf{202}, 06001 (2019)}.

\bibitem{Faustov:2021qqf}
R.~N.~Faustov and V.~O.~Galkin,
``Triply heavy baryon spectroscopy in the relativistic quark model,''
\href{https://doi:10.1103/PhysRevD.105.014013}{Phys. Rev. D \textbf{105}, 014013 (2022)}.

\bibitem{Flynn:2011gf}
J.~M.~Flynn, E.~Hernandez and J.~Nieves,
``Triply Heavy Baryons and Heavy Quark Spin Symmetry,''
\href{https://doi:10.1103/PhysRevD.85.014012}{Phys. Rev. D \textbf{85}, 014012 (2012)}.

\bibitem{Vijande:2004at}
J.~Vijande, H.~Garcilazo, A.~Valcarce and F.~Fernandez,
``Spectroscopy of doubly charmed baryons,''
\href{https://doi:10.1103/PhysRevD.70.054022}{Phys. Rev. D \textbf{70}, 054022 (2004)}.

\bibitem{Wang:2011ae}
Z.~G.~Wang,
``Analysis of the Triply Heavy Baryon States with QCD Sum Rules,''
\href{https://doi:10.1088/0253-6102/58/5/17}{Commun. Theor. Phys. \textbf{58}, 723 (2012)}.

\bibitem{Wang:2019gal}
Z.~G.~Wang,
``Triply-charmed dibaryon states or two-baryon scattering states from QCD sum rules,''
\href{https://doi:10.1103/PhysRevD.102.034008}{Phys. Rev. D \textbf{102}, 034008 (2020)}.

\bibitem{Wang:2020avt}
Z.~G.~Wang,
``Analysis of the triply-heavy baryon states with the QCD sum rules,''
\href{https://doi:10.1007/s43673-021-00006-3}{AAPPS Bull. \textbf{31}, 5 (2021)}.

\bibitem{Aliev:2012tt}
T.~M.~Aliev, K.~Azizi and M.~Savci,
``Masses and Residues of the Triply Heavy Spin-1/2 Baryons,''
\href{https://doi:10.1007/JHEP04(2013)042}{JHEP \textbf{04}, 042 (2013)}.

\bibitem{Azizi:2014jxa}
K.~Azizi, T.~M.~Aliev and M.~Savci,
``Properties of doubly and triply heavy baryons,''
\href{https://doi:10.1088/1742-6596/556/1/012016}{J. Phys. Conf. Ser. \textbf{556}, 012016 (2014)}.

\bibitem{Zhang:2009re}
J.~R.~Zhang and M.~Q.~Huang,
``Deciphering triply heavy baryons in terms of QCD sum rules,''
\href{https://doi:10.1016/j.physletb.2009.02.056}{Phys. Lett. B \textbf{674}, 28 (2009)}.

\bibitem{Aliev:2014lxa}
T.~M.~Aliev, K.~Azizi and M.~Savc\i{},
``Properties of triply heavy spin-3/2 baryons,''
\href{https://doi:10.1088/0954-3899/41/6/065003}{J. Phys. G \textbf{41}, 065003 (2014)}.

\bibitem{Meinel:2010pw}
S.~Meinel,
``Prediction of the $\Omega_{bbb}$ mass from lattice QCD,''
\href{https://doi:10.1103/PhysRevD.82.114514}{Phys. Rev. D \textbf{82}, 114514 (2010)}.

\bibitem{Meinel:2012qz}
S.~Meinel,
``Excited-state spectroscopy of triply-bottom baryons from lattice QCD,''
\href{https://doi:10.1103/PhysRevD.85.114510}{Phys. Rev. D \textbf{85}, 114510 (2012)}.

\bibitem{Padmanath:2013zfa}
M.~Padmanath, R.~G.~Edwards, N.~Mathur and M.~Peardon,
``Spectroscopy of triply-charmed baryons from lattice QCD,''
\href{https://doi:10.1103/PhysRevD.90.074504}{Phys. Rev. D \textbf{90}, 074504 (2014)}.

\bibitem{PACS-CS:2013vie}
Y.~Namekawa \textit{et al.} [PACS-CS],
``Charmed baryons at the physical point in 2+1 flavor lattice QCD,''
\href{https://doi:10.1103/PhysRevD.87.094512}{Phys. Rev. D \textbf{87}, 094512 (2013)}.

\bibitem{Vijande:2015faa}
J.~Vijande, A.~Valcarce and H.~Garcilazo,
``Constituent-quark model description of triply heavy baryon nonperturbative lattice QCD data,''
\href{https://doi:10.1103/PhysRevD.91.054011}{Phys. Rev. D \textbf{91}, 054011 (2015)}.

\bibitem{Mathur:2018epb}
N.~Mathur, M.~Padmanath and S.~Mondal,
``Precise predictions of charmed-bottom hadrons from lattice QCD,''
\href{https://doi:10.1103/PhysRevLett.121.202002}{Phys. Rev. Lett. \textbf{121}, 202002 (2018)}.


\bibitem{Can:2015exa}
K.~U.~Can, G.~Erkol, M.~Oka and T.~T.~Takahashi,
``Look inside charmed-strange baryons from lattice QCD,''
\href{https://doi:10.1103/PhysRevD.92.114515}{Phys. Rev. D \textbf{92}, 114515 (2015)}.

\bibitem{Brown:2014ena}
Z.~S.~Brown, W.~Detmold, S.~Meinel and K.~Orginos,
``Charmed bottom baryon spectroscopy from lattice QCD,''
\href{https://doi:10.1103/PhysRevD.90.094507}{Phys. Rev. D \textbf{90}, 094507 (2014)}.

\bibitem{Briceno:2012wt}
R.~A.~Briceno, H.~W.~Lin and D.~R.~Bolton,
``Charmed-Baryon Spectroscopy from Lattice QCD with $N_f$ = 2+1+1 Flavors,''
\href{https://doi:10.1103/PhysRevD.86.094504}{Phys. Rev. D \textbf{86}, 094504 (2012)}.

\bibitem{Wei:2015gsa}
K.~W.~Wei, B.~Chen and X.~H.~Guo,
``Masses of doubly and triply charmed baryons,''
\href{https://doi:10.1103/PhysRevD.92.076008}{Phys. Rev. D \textbf{92}, 076008 (2015)}.

\bibitem{Wei:2016jyk}
K.~W.~Wei, B.~Chen, N.~Liu, Q.~Q.~Wang and X.~H.~Guo,
``Spectroscopy of singly, doubly, and triply bottom baryons,''
\href{https://doi:10.1103/PhysRevD.95.116005}{Phys. Rev. D \textbf{95}, 116005 (2017)}.

\bibitem{Oudichhya:2023pkg}
J.~Oudichhya, K.~Gandhi and A.~k.~Rai,
``Investigation of $\Omega _{ccb}$ and $\Omega _{cbb}$ baryons in Regge phenomenology,''
\href{https://doi:10.1007/s12043-023-02630-0}{Pramana \textbf{97}, 151 (2023)}.

\bibitem{Brambilla:2009cd}
N.~Brambilla, J.~Ghiglieri and A.~Vairo,
``The Three-quark static potential in perturbation theory,''
\href{https://doi:10.1103/PhysRevD.81.054031}{Phys. Rev. D \textbf{81}, 054031 (2010)}.

\bibitem{Llanes-Estrada:2011gwu}
F.~J.~Llanes-Estrada, O.~I.~Pavlova and R.~Williams,
``A First Estimate of Triply Heavy Baryon Masses from the pNRQCD Perturbative Static Potential,''
\href{https://doi:10.1140/epjc/s10052-012-2019-9}{Eur. Phys. J. C \textbf{72}, 2019 (2012)}.

\bibitem{Gutierrez-Guerrero:2019uwa}
L.~X.~Guti\'errez-Guerrero, A.~Bashir, M.~A.~Bedolla and E.~Santopinto,
``Masses of Light and Heavy Mesons and Baryons: A Unified Picture,''
\href{https://doi:10.1103/PhysRevD.100.114032}{Phys. Rev. D \textbf{100}, 114032 (2019)}.

\bibitem{Brambilla:2005yk}
N.~Brambilla, A.~Vairo and T.~Rosch,
``Effective field theory Lagrangians for baryons with two and three heavy quarks,''
\href{https://doi:10.1103/PhysRevD.72.034021}{Phys. Rev. D \textbf{72}, 034021 (2005)}.

\bibitem{Yin:2019bxe}
P.~L.~Yin, C.~Chen, G.~Krein, C.~D.~Roberts, J.~Segovia and S.~S.~Xu,
``Masses of ground-state mesons and baryons, including those with heavy quarks,''
\href{https://doi:10.1103/PhysRevD.100.034008}{Phys. Rev. D \textbf{100}, 034008 (2019)}.

\bibitem{Qin:2019hgk}
S.~x.~Qin, C.~D.~Roberts and S.~M.~Schmidt,
``Spectrum of light- and heavy-baryons,''
\href{https://doi:10.1007/s00601-019-1488-x}{Few Body Syst. \textbf{60}, 26 (2019)}.

\bibitem{Serafin:2018aih}
K.~Serafin, M.~G\'omez-Rocha, J.~More and S.~D.~G\l{}azek,
``Approximate Hamiltonian for baryons in heavy-flavor QCD,''
\href{https://doi:10.1140/epjc/s10052-018-6436-2}{Eur. Phys. J. C \textbf{78}, 964 (2018)}.

\bibitem{Ebert:2002pp}
D.~Ebert, R.~N.~Faustov and V.~O.~Galkin,
``Properties of heavy quarkonia and $B_c$ mesons in the relativistic quark model,''
\href{https://doi:10.1103/PhysRevD.67.014027}{Phys. Rev. D \textbf{67}, 014027 (2003)}.

\bibitem{Ebert:2011jc}
D.~Ebert, R.~N.~Faustov and V.~O.~Galkin,
``Spectroscopy and Regge trajectories of heavy quarkonia and $B_c$ mesons,''
\href{https://doi:10.1140/epjc/s10052-011-1825-9}{Eur. Phys. J. C \textbf{71}, 1825 (2011)}.

\bibitem{GI1}
S.~Godfrey and N.~Isgur,
``Mesons in a Relativized Quark Model with Chromodynamics,''
\href{https://doi.org/10.1103/PhysRevD.32.189}{Phys. Rev. D \textbf{32}, 189 (1985)}.

\bibitem{GI2}
S.~Capstick and N.~Isgur,
``Baryons in a relativized quark model with chromodynamics,''
\href{https://doi.org/10.1103/physrevd.34.2809}{Phys. Rev. D \textbf{34}, 2809 (1986)};
\href{https://doi.org/10.1063/1.35361}{AIP Conf. Proc. \textbf{132}, 267 (1985)}.

\bibitem{LV1}
Q.~F.~L\"u, D.~Y.~Chen and Y.~B.~Dong,
``Masses of fully heavy tetraquarks $QQ {\bar{Q}} {\bar{Q}}$ in an extended relativized quark model,''
\href{https://doi.org/10.1140/epjc/s10052-020-08454-1}{Eur. Phys. J. C \textbf{80}, 871 (2020)}.

\bibitem{LV2}
Q.~F.~L\"u, D.~Y.~Chen, Y.~B.~Dong and E.~Santopinto,
``Triply-heavy tetraquarks in an extended relativized quark model,''
\href{https://doi.org/10.1103/PhysRevD.104.054026}{Phys. Rev. D \textbf{104}, 054026 (2021)}.

\bibitem{Wang:2021kfv}
G.~J.~Wang, L.~Meng, M.~Oka and S.~L.~Zhu,
``Higher fully charmed tetraquarks: Radial excitations and P-wave states,''
\href{https://doi:10.1103/PhysRevD.104.036016}{Phys. Rev. D \textbf{104}, 036016 (2021)}.

\bibitem{Liu:2020lpw}
F.~X.~Liu, M.~S.~Liu, X.~H.~Zhong and Q.~Zhao,
``Fully-strange tetraquark $ss\bar{s}\bar{s}$ spectrum and possible experimental evidence,''
\href{https://doi:10.1103/PhysRevD.103.016016}{Phys. Rev. D \textbf{103}, 016016 (2021)}.

\bibitem{Meng:2023jqk}
L.~Meng, Y.~K.~Chen, Y.~Ma and S.~L.~Zhu,
``Tetraquark bound states in constituent quark models: Benchmark test calculations,''
\href{https://doi:10.1103/PhysRevD.108.114016}{Phys. Rev. D \textbf{108}, 114016 (2023)}.

\bibitem{Yu:2022lak}
G.~L.~Yu, Z.~Y.~Li, Z.~G.~Wang, J.~Lu and M.~Yan,
``The S- and P-wave fully charmed tetraquark states and their radial excitations,''
\href{https://doi:10.1140/epjc/s10052-023-11445-7}{Eur. Phys. J. C \textbf{83}, 416 (2023)}.

\bibitem{Yu:2024ljg}
G.~L.~Yu, Z.~Y.~Li, Z.~G.~Wang, B.~WU, Z.~Zhou and J.~Lu,
``The ground states of hidden-charm tetraquarks and their radial excitations,''
\href{https://doi:10.1140/epjc/s10052-024-13514-x}{Eur. Phys. J. C \textbf{84}, 1130 (2024)}.

\bibitem{GLY1}
G.~L.~Yu, Z.~Y.~Li, Z.~G.~Wang, J.~Lu and M.~Yan,
``Systematic analysis of single heavy baryons $\Lambda_{Q}$, $\Sigma_{Q}$ and $\Omega_{Q}$,''
\href{https://doi.org/10.1016/j.nuclphysb.2023.116183}{Nucl. Phys. B \textbf{990}, 116183 (2023)}.

\bibitem{Yu:2022lel}
G.~L.~Yu, Z.~Y.~Li, Z.~G.~Wang, J.~Lu and M.~Yan,
``Systematic analysis of doubly charmed baryons $\Xi _{cc}$ and $\Omega _{cc}$,''
\href{https://doi:10.1140/epja/s10050-023-01044-1}{Eur. Phys. J. A \textbf{59},126 (2023)}.

\bibitem{ZYL1}
Z.~Y.~Li, G.~L.~Yu, Z.~G.~Wang, J.~Z.~Gu and J.~Lu,
``Systematic analysis of strange single heavy baryons $\Xi_{c}$ and $\Xi_{b}$,''
\href{https://doi:10.1088/1674-1137/acd365}{Chin. Phys. C \textbf{47}, 073105 (2023)}.

\bibitem{Li:2023gbo}
Z.~Y.~Li, G.~L.~Yu, Z.~G.~Wang and J.~Z.~Gu,
``Heavy quark dominance in orbital excitation of singly and doubly heavy baryons,''
\href{https://doi:10.1140/epjc/s10052-024-12457-7}{Eur. Phys. J. C \textbf{84}, no.2, 106 (2024)}.

\bibitem{LHCb:2023sxp}
R.~Aaij \textit{et al.} [LHCb],
``Observation of New $\Omega_{c}^{0}$ States Decaying to the $\Xi_{c}^{+}K^{-}$ Final State,''
\href{https://doi:10.1103/PhysRevLett.131.131902}{Phys. Rev. Lett. \textbf{131}, 131902 (2023)}.

\bibitem{LHCb:2023zpu}
R.~Aaij \textit{et al.} [LHCb],
``Observation of New Baryons in the $\Xi_{b}^{-}\pi^{+}\pi^{-}$ and $\Xi_{b}^{0}\pi^{+}\pi^{-}$ Systems,''
\href{https://doi:10.1103/PhysRevLett.131.171901}{Phys. Rev. Lett. \textbf{131}, 171901 (2023)}.

\bibitem{Gaussian1}
M.~Kamimura,
``Nonadiabatic coupled-rearrangement-channel approach to muonic molecules,''
\href{https://doi.org/10.1103/PhysRevA.38.621}{Phys. Rev. A \textbf{38}, 621 (1988)}.

\bibitem{Gaussian2}
E.~Hiyama, Y.~Kino and M.~Kamimura,
``Gaussian expansion method for few-body systems,''
\href{https://doi.org/10.1016/S0146-6410(03)90015-9}{Prog. Part. Nucl. Phys. \textbf{51}, 223 (2003)}.

\bibitem{Gaussian3}
Y.~W.~Pan, T.~W.~Wu, M.~Z.~Liu and L.~S.~Geng,
``Three-body molecules $\overline{D} \overline{D}^{*}\Sigma_{C}$: Understanding the nature of $T_{cc}$, $P_c$(4312), $P_c$(4440), and $P_c$(4457),''
\href{https://doi.org/10.1103/PhysRevD.105.114048}{Phys. Rev. D \textbf{105}, 114048 (2022)[arXiv:2204.02295 [hep-ph]]}.

\bibitem{Regge1}
T.~Regge,
``Introduction to complex orbital momenta,''
\href{https://doi.org/10.1007/BF02728177}{Nuovo Cim. \textbf{14}, 951 (1959)}.

\bibitem{Regge2}
T.~Regge,
``Bound states, shadow states and Mandelstam representation,''
\href{https://doi.org/10.1007/BF02733035}{Nuovo Cim. \textbf{18}, 947 (1960)}.

\bibitem{Regge3}
G.~F.~Chew and S.~C.~Frautschi,
``Principle of Equivalence for All Strongly Interacting Particles Within the S Matrix Framework,''
\href{https://doi.org/10.1103/PhysRevLett.7.394}{Phys. Rev. Lett. \textbf{7}, 394 (1961)}.

\bibitem{Regge4}
G.~F.~Chew and S.~C.~Frautschi,
``Regge Trajectories and the Principle of Maximum Strength for Strong Interactions,''
\href{https://doi.org/10.1103/PhysRevLett.8.41}{Phys. Rev. Lett. \textbf{8}, 41 (1962)}.

\bibitem{Regge5}
G.~S.~Bali,
``QCD forces and heavy quark bound states,''
\href{https://doi.org/10.1016/S0370-1573(00)00079-X}{Phys. Rept. \textbf{343}, 1 (2001)}.

\bibitem{Regge6}
D.~V.~Bugg,
``Four sorts of meson,''
\href{https://doi.org/10.1016/j.physrep.2004.03.008}{Phys. Rept. \textbf{397}, 257 (2004)}.

\bibitem{Regge7}
E.~Klempt and A.~Zaitsev,
``Glueballs, Hybrids, Multiquarks. Experimental facts versus QCD inspired concepts,''
\href{https://doi.org/10.1016/j.physrep.2007.07.006}{Phys. Rept. \textbf{454}, 1 (2007)}.

\bibitem{Regge8}
W.~Lucha, F.~F.~Schoberl and D.~Gromes,
``Bound states of quarks,''
\href{https://doi.org/10.1016/0370-1573(91)90001-3}{Phys. Rept. \textbf{200}, 127 (1991)}.

\bibitem{Regge9}
Y.~Nambu,
``Strings, Monopoles and Gauge Fields,''
\href{https://doi.org/10.1103/PhysRevD.10.4262}{Phys. Rev. D \textbf{10}, 4262 (1974)}.

\bibitem{Regge10}
Y.~Nambu,
``QCD and the String Model,''
\href{https://doi.org/10.1016/0370-2693(79)91193-6}{Phys. Lett. B \textbf{80}, 372 (1979)}.

\bibitem{Ebert}
D.~Ebert, R.~N.~Faustov and V.~O.~Galkin,
``Spectroscopy and Regge trajectories of heavy baryons in the relativistic quark-diquark picture,''
\href{https://doi.org/10.1103/PhysRevD.84.014025}{Phys. Rev. D \textbf{84}, 014025 (2011)}.

\bibitem{Guo:2008he}
X.~H.~Guo, K.~W.~Wei and X.~H.~Wu,
``Some mass relations for mesons and baryons in Regge phenomenology,''
\href{https://doi:10.1103/PhysRevD.78.056005}{Phys. Rev. D \textbf{78}, 056005 (2008)}.
\end{thebibliography}
\end{document}